\documentclass[preprintnumbers, floatfix, letterpaper, onecolumn,aps,prd,epsfig,nofootinbib,natbib,
longbibliography
]{revtex4-2}
\usepackage{bm,graphicx,dcolumn,epstopdf,epsf, latexsym,mathbbol, amssymb,amsmath,color,slashed, mathrsfs,mathcomp,simplewick}
\usepackage[center]{subfigure}
\usepackage{multirow}
\usepackage{graphicx}
\usepackage{makecell}
\usepackage{epstopdf}
\usepackage[table]{xcolor}
\usepackage[colorlinks,linkcolor=blue,citecolor=blue,urlcolor=blue]{hyperref}

\begin{document}
\allowdisplaybreaks
 \newcommand{\bq}{\begin{equation}}
 \newcommand{\eq}{\end{equation}}
 \newcommand{\bqn}{\begin{eqnarray}}
 \newcommand{\eqn}{\end{eqnarray}}
 \newcommand{\nb}{\nonumber}
 \newcommand{\lb}{\label}
\newcommand{\f}{\frac}
\newcommand{\p}{\partial}
\newcommand{\PRL}{Phys. Rev. Lett.}
\newcommand{\PLB}{Phys. Lett. B}
\newcommand{\PRD}{Phys. Rev. D}
\newcommand{\CQG}{Class. Quantum Grav.}
\newcommand{\JCAP}{J. Cosmol. Astropart. Phys.}
\newcommand{\JHEP}{J. High. Energy. Phys.}
\newcommand{\orcid}[1]{\href{https://orcid.org/#1}{\includegraphics[width=10pt]{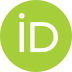}}}

\title{
  Inflationary Dynamics of Mutated Hilltop Inflation in Einstein-Gauss-Bonnet Gravity Under New Slow-Roll Approximations with Generalised Reheating}
\author{Yogesh\orcid{0000-0002-7638-3082}}
\email{yogeshjjmi@gmail.com}
\affiliation{Institute for Theoretical Physics and Cosmology,
Zhejiang University of Technology, Hangzhou, 310023, China}

\author{Mehnaz Zahoor\orcid{0009-0009-7926-4861}}
\email{mehnazzahoormir@gmail.com}
\affiliation{Department of Physics, Central University of Kashmir, Ganderbal J\&K- 191131 }
\author{Kashif Ali Wani\orcid{0009-0001-6546-0265} }
\email{kashifaliwani@gmail.com}
\affiliation{Department of Physics, Central University of Kashmir, Ganderbal J\&K- 191131 }

\author{Imtiyaz Ahmad Bhat\orcid{0000-0002-2695-9709}}
\email{imtiyaz@ctp-jamia.res.in}
\affiliation{Department of Physics, Islamic University of Science and Technology, Awantipora, J\&K- 192122 }


\begin{abstract}
The advancement in the observational cosmology of the early universe such as Cosmic Microwave Background (CMB) observations, puts severe constraints on the inflationary models. Many inflationary models have been ruled out by CMB, nevertheless the models ruled out in standard cold inflationary scenarios can be resurrected in modified gravity models. In this regard we examine the dynamics of inflation within the framework of Einstein-Gauss-Bonnet (EGB) Gravity using the new slow-roll approximation methods proposed in  \cite{Pozdeeva:2024ihc}. We consider the Mutated Hilltop inflation model~\cite{Pal:2009sd,Pinhero:2019sbg} due to its origin from super-gravity, a naturally perfect choice to study the impact of EGB on inflationary observables such as tensor-to-scalar ratio ($r$) and scalar spectral index ($n_s$). The period of reheating following the inflationary phase is also examined, and for the {\it Planck'18} permitted values of $n_s$, constraints on the reheating temperature ($T_{re}$) are computed for various equations of states during reheating ($\omega_{re}$). 
\end{abstract}

    \maketitle

\vspace{0.0001in}


\baselineskip=15.4pt

\section{Introduction\label{sec:intro}}
\renewcommand{\theequation}{1.\arabic{equation}} \setcounter{equation}{0}
The current state of the universe cannot be explained by the standard hot Big Bang model alone. To rectify its shortcomings a phase of inflation is needed \cite{Liddle:2000cg,1990eaun.book.....K}. Since Guth \cite{Guth:1980zm} initiated the idea of inflation, inflationary cosmology has drawn a great deal of attention and developed into an interdisciplinary area requiring knowledge of both cosmology and high energy physics \cite{Linde:1981mu,Mukhanov:1981xt,Sato:1981qmu,1996tyli.conf..771S,PhysRevLett.48.1220,Starobinsky:1982ee}. A plethora of inflationary models have been ruled out by Wilkinson Microwave Anisotropy Probe (WMAP) and CMB observations by the {\it Planck}~\cite{Hinshaw_2013, Planck:2015sxf,Planck:2018jri} in the standard cold inflationary scenario. It has been studied that the  inflationary models that are in conflict with observation in the standard cold inflationary framework, can nonetheless satisfy the observational constraints in the modified gravity framework~\cite{Kallosh:2013pby,1982PhLB..117..175S,Starobinsky:1983zz,Barvinsky:1994hx,Cervantes-Cota:1995ehs,Bezrukov:2007ep,Barvinsky:2008ia,DeSimone:2008ei,Gialamas:2020vto,Bezrukov:2008ej,Barvinsky:2009ii,Bezrukov:2010jz,Bezrukov:2013fka,Rubio:2018ogq,Koshelev:2020xby,Elizalde:2014xva,Gialamas:2023flv,Gialamas:2022xtt,Gialamas:2021enw,Gialamas:2020snr}.
In this paper, we investigate inflation in the framework of  EGB gravity, a higher-dimensional theory that modifies General Relativity by quadratic curvature corrections. When we introduce a coupling between a scalar field and the Gauss-Bonnet term, EGB gravity provides a rich ambiance for studying non-standard inflationary dynamics.
In EGB the total derivative of the equation of motion remains invariant after adding the Gauss-Bonnet component to the Einstein-Hilbert action. EGB coupling $\xi(\phi)$ becomes dynamically essential when paired with a scalar field $\phi$, the Gauss-Bonnet term acts as a quantum correction to the Einstein-Hilbert action in string theory. A large number of inflationary models have been studied in the modified gravity framework~\cite{vandeBruck:2015gjd,Guo:2009uk,Guo:2010jr,Koh:2016abf,Pozdeeva:2020shl,Satoh:2008ck,Jiang:2013gza,Koh:2014bka,Koh:2018qcy,Mathew:2016anx,Mathew:2016anx,Pozdeeva:2020apf,Pozdeeva:2016cja,Nozari:2017rta,Armaleo:2017lgr,Yi:2018gse,Yi:2018dhl,Odintsov:2018zhw,Fomin:2019yls,Fomin:2020hfh,Kleidis:2019ywv,Rashidi:2020wwg,Odintsov:2020sqy,Odintsov:2020zkl,Kawai:2021bye,Kawai:2017kqt,Oikonomou:2022xoq,Oikonomou:2022ksx,Cognola:2006sp,Odintsov:2020xji,Odintsov:2020mkz,Oikonomou:2020sij,Nojiri:2019dwl,Fomin:2019yls,Ashrafzadeh:2024oll,Solbi:2024zhl,Ashrafzadeh:2023ndt,Oikonomou:2024etl,Oikonomou:2024jqv,Odintsov:2023weg,Kawai:2023nqs,Kawai:1999pw,Kawai:1998ab,Nojiri:2024hau,Nojiri:2024zab,Elizalde:2023rds,Nojiri:2023mvi,Odintsov:2023aaw,Odintsov:2022rok,Odintsov:2022rok,Odintsov:2021urx}. In most of the cases the EGB coupling studied was the inverse function of the scalar field~\cite{Guo:2009uk,Guo:2010jr,Koh:2016abf,Pozdeeva:2020shl,Jiang:2013gza,Yi:2018gse,Odintsov:2018zhw,Kleidis:2019ywv,Rashidi:2020wwg}. 
  However, in this work, we will consider a different form of coupling involving a $``\tanh"$ term~\cite{Kawai:2021edk}. Recently in~\cite{Pozdeeva:2024ihc}, the authors have presented a new form of slow-roll approximations, which are found to be more accurate in comparison to the previous slow-roll approximations presented in \cite{Pozdeeva:2020apf}. As it has already been shown in \cite{Pozdeeva:2024ihc} that new slow-roll approximations are more precise, we aim to do a case study of the new slow-roll approximations in the EGB scenario using the Mutated Hilltop inflation. Mutated  Hilltop inflation is a well-motivated model that emerges from the super-gravity~\cite{Pal:2009sd,Pinhero:2019sbg}.  Moreover, we also examine the generalized reheating and study its effect on the effective EGB background. The standard cold inflationary scenario demands a reheating phase after the termination of inflation to metastasize into the radiation-dominated era, thereby initiating Big Bang Nucleosynthesis (BBN). The conceptual model building of the inflationary models relies on satisfying the constraints of the recent observations and reheating. In this article, we have investigated the reheating and shown that the expansion of the universe during the reheating(e-folds during reheating) epoch can be constrained through CMB observations.

 The rest of the paper is organized as follows: In section~\ref{EGB}, we will go over some
of the fundamental critical features in the EGB framework. In section~\ref{sraw},  we review the slow-roll parameters without any approximations. In section~\ref{sranew}, the new slow-roll approximations (SRA) as reported and mentioned in~\cite{Pozdeeva:2024ihc} are reviewed. We report our results for the Mutated Hilltop under the new slow-roll approximations along with the numerical calculation and generalized reheating in the sections \ref{infa} and \ref{sec:reh} respectively. Lastly in section~\ref{discussions}, we conclude with our conclusions and future direction of work.
\section{Review of Einstein-Gauss Bonnet Gravity}
\renewcommand{\theequation}{2.\arabic{equation}} \setcounter{equation}{0}
\label{EGB}
 In the Einstein-Gauss-Bonnet gravity framework, the effective action can be written as: 

\begin{equation}
\label{action1}
S=\int d^4x\sqrt{-g}\left[U_0R-\frac{1}{2}g^{\mu\nu}\partial_\mu\phi\partial_\nu\phi-V(\phi)-\frac{1}{2}\xi(\phi){\cal G}\right],
\end{equation}
where $U_0>0$ is a constant, $V(\phi)$ and $\xi(\phi)$ are the potential and EGB coupling respectively, which are functions of field $\phi$. Also, $R$ stands for the usual Ricci scalar. The EGB term is given by: 
\begin{equation}
   \mathcal{G}=R_{\mu\nu\rho\sigma}R^{\mu\nu\rho\sigma}-4R_{\mu\nu}R^{\mu\nu}+R^2. 
\end{equation}
The dynamical system and evolution equations for the spatially flat FLRW metric in Einstein-Gauss-Bonnet gravity can be expressed \cite{vandeBruck:2015gjd,Pozdeeva:2020apf} as:
\begin{eqnarray}
12H^2\left(U_0-2\xi_{,\phi}\psi H\right)&=&\psi^2+2V, 
\label{Equ00} \\
4\dot H\left(U_0-2\xi_{,\phi}\psi H\right)&=&{}-\psi^2+4H^2\left(\xi_{,\phi\phi}\psi^2+\xi_{,\phi}\dot\psi-H\xi_{,\phi}\psi\right), 
\label{Equ11} \\
\dot{\psi}+3H\psi&=&{} -V_{,\phi} -12H^2\xi_{,\phi}\left(\dot{H}+H^2\right),\label{Equphi}
\end{eqnarray}
where $H=\dot{a}/a$ signifies the Hubble parameter and  $a(t)$ denotes the scale factor. Here, the dot represents the time derivative. The term $\psi=\dot{\phi}$ and $A_{,\phi} \equiv \frac{dA}{d\phi}$ (true for any function $A(\phi)$).
Considering the time derivative of Eq.~(\ref{Equ00}), we get
\begin{equation}
\label{Dequ00}
12H\dot{H}\left(U_0-2\xi_{,\phi}H\psi\right)-12H^2\left(\xi_{,\phi\phi}H{\psi}^2+\xi_{,\phi}\left[H\dot{\psi}+\dot{H}\psi\right]\right)-\psi\dot{\psi}-V_{,\phi}\psi=0.
\end{equation}

    Expressing $V_{,\phi}$ from (\ref{Dequ00}):
\begin{equation}
\label{VphiDequ00}
V_{,\phi}=12H\dot{H}\left(\frac{U_0}{\psi}-2\xi_{,\phi}H\right)-12H^2\left(\xi_{,\phi\phi} H\psi+\xi_{,\phi}H\frac{\dot\psi}{\psi}+\xi_{,\phi}\dot{H}\right)-\dot\psi\,,
\end{equation}
and substituting it to (\ref{Equphi}), we get  Eq.~(\ref{Equ11}) is a result of Eqs.~(\ref{Equ00}) and (\ref{Equphi}).

 When $\xi_{,\phi}\neq 0$, then equations (\ref{Equ11}) and (\ref{Equphi}) do not form a dynamic system. Instead, it is efficient to use Eqs.~(\ref{Equ11}) and (\ref{Equphi}) to obtain a dynamical system~\cite{Pozdeeva:2019agu}:
\begin{equation}
\label{DynSYS}
\begin{split}
\dot\phi=&\psi,\\~~~~~~~~~~~~~
\dot\psi=&\frac{1}{B-2\xi_{,\phi}H\psi}\left\{
3\left[3-4\,\xi_{,\phi\phi} H^2 \right]\xi_{,\phi}H^2\psi^2+\left[3B+2\,\xi_{,\phi}V_{,\phi}-6U_0\right]H\psi-\frac{V^2}{U_0}X\right\},\\
\dot H=&\frac{1}{4\left(B-2\,\xi_{,\phi}H\psi\right)}\left\{\left(4\,\xi_{,\phi\phi}H^2-1\right)\psi^2-16\xi_{,\phi}H^3\psi-4\frac{V^2}{U_0^2}\,\xi_{,\phi} H^2 X\right\},
\end{split}
\end{equation}\\
where
\begin{equation}
B=12\xi_{,\phi}^2H^4+U_0,\qquad
X=\frac{U_0^2}{V^2}\left(12\xi_{,\phi} H^4+V_{,\phi} \right).
\end{equation}

       Allowing $\dot{\psi}$ from Eq.~(\ref{Equphi}) to substitute into Eq.~(\ref{Dequ00}), we find the third equation of system (\ref{DynSYS}). So, this equation is a consequence of Eqs.~(\ref{Equ00}) and (\ref{Equphi}).
For an inflationary model, the duration of the inflation can be denoted in terms of number of e-folds, which is given by $N=\ln(a/a_{e})$. Admitting ${d}/{dt}=H\, {d}/{dN}$ and introducing $\chi=\psi/H$, one can express the system~(\ref{DynSYS}) as follows:
    
\begin{equation}
\label{DynSYSN}
\begin{split}
\frac{d\phi}{dN}=&\,\chi,\\
\frac{d\chi}{dN}=&\,\frac{1}{Q\left(B-2\xi_{,\phi}Q\chi\right)}\left\{
3\left[3-4\xi_{,\phi\phi} Q\right]\xi_{,\phi}Q^2\chi^2+ \left[3B+2\xi_{,\phi}V_{,\phi}-6U_0\right]Q\chi-\frac{V^2}{U_0}X\right\}\\
&{}-\frac{\chi}{2Q}\frac{dQ}{dN},\\
\frac{dQ}{dN}=&\,\frac{Q}{2\left(B-2\xi_{,\phi}Q\chi\right)}\left\{\left(4\xi_{,\phi\phi}Q-1\right)\chi^2-16\xi_{,\phi}Q\chi-4\frac{V^2}{U_0^2}\xi_{,\phi}X\right\},
\end{split}
\end{equation}

where $Q\equiv H^2$ and rewriting the Eq. (\ref{Equ00}) in the following form:
\begin{equation}
\label{Equ00N}
24\xi_{,\phi}\chi Q^2+\left(\chi^2- 12U_0\right)Q+2V=0.
\end{equation}
and has solutions:
\begin{equation}
\label{Q0}
Q_\pm=\frac{12U_0-\chi^2\pm\sqrt{\left(12U_0-\chi^2\right)^2-192V\xi_{,\phi}\chi}}{48\xi_{,\phi}\chi},
\end{equation}

if $\xi_{,\phi}\chi\neq 0$, then, a unique solution will exist in the opposite case.
\begin{equation}
Q_0=\frac{2V}{12U_0-\chi^2}\,.
\end{equation}
Equation (\ref{Equ00N}) impose the restricts on initial conditions of system~(\ref{DynSYSN}).

    Transforming equations(\ref{Equ11}) and (\ref{Equphi}),we get:
\begin{equation}
 2\frac{dQ}{dN}\left(U_0-2\xi_{,\phi}\chi Q\right)={}-Q\chi^2+ 4Q\left(\xi_{,\phi\phi}Q\chi^2+ \xi_{,\phi}\left( \frac{\chi}{2}\frac{dQ}{dN}+Q\frac{d\chi}{dN}\right)-\xi_{,\phi}Q\chi\right), 
\label{Equ11N}    
\end{equation}              
\begin{equation}
\frac{\chi}{2}\,\frac{dQ}{dN}+Q\,\frac{d\chi}{dN}+3Q\,\chi={} -V_{,\phi} -6Q\,\xi_{,\phi}\left(\frac{dQ}{dN}+2Q\right).
\label{EquphiN}
\end{equation}   



\section{The slow-roll parameters without any approximation}
\renewcommand{\theequation}{3.\arabic{equation}} \setcounter{equation}{0}
\label{sraw}
~~~~Following Refs.~\cite{Guo:2010jr,vandeBruck:2015gjd,Pozdeeva:2020apf,Odintsov:2023lbb}, we consider the slow-roll parameters:
\begin{equation}
\label{epsilon}
\varepsilon_1 ={}-\frac{\dot{H}}{H^2}={}-\frac{d\ln(H)}{dN},\qquad \varepsilon_{i+1}= \frac{d\ln|\varepsilon_i|}{dN},\quad i\geqslant 1,
\end{equation}
\begin{equation}
\label{delta}
\delta_1= \frac{2}{U_0}\xi_{,\phi}H\psi=\frac{2}{U_0}\xi_{,\phi}H^2\chi,\qquad \delta_{i+1}=\frac{d\ln|\delta_i|}{dN}, \quad i\geqslant 1.
\end{equation}

We can easily see that
\begin{equation}
\label{delta2}
\delta_2=\frac{\dot{\psi}}{H\psi}+\frac{\xi_{,\phi\phi}\psi}{H\xi_{,\phi}}-\varepsilon_1.
\end{equation}
 
   Now, using system~(\ref{DynSYSN}), one can obtain, that the parameter $\varepsilon_1(N)$ satisfies the following equation:
\begin{equation}
\label{equepsilon1a}
\varepsilon_1={}-\frac{1}{2Q}\frac{dQ}{dN}={}-\frac{1}{4\left(B-2\xi_{,\phi}Q\chi\right)}\left\{\left(4\xi_{,\phi\phi}Q-1\right)\chi^2-16\xi_{,\phi}Q\chi-4\frac{V^2}{U_0^2}\xi_{,\phi}X\right\}\,.
\end{equation}

    Additionally, another form of $\varepsilon_1$ can be computed by dividing Eq.~(\ref{Equ11}) on Eq.~(\ref{Equ00}),
\begin{equation}
\label{equepsilon1b}
\varepsilon_1=\frac{3}{\psi^2+2V}\left[\psi^2-4H^2\left(\xi_{,\phi\phi}\psi^2+\xi_{,\phi}\dot\psi-H\xi_{,\phi}\psi\right)\right].
\end{equation}

 It is worth to mentioned that the expression for $\varepsilon_1$ obtained here is exact without employing any approximation.

    It is advantageous to redraft evolution equations in terms of the slow-roll parameters.
Equations~(\ref{Equ00}) and (\ref{Equ11}) are equivalent to
\begin{equation}
\label{Equ00delta1}
12U_0H^2\left(1-\delta_1\right)=\psi^2+2V=H^2\chi^2+2V\,,
\end{equation}
\begin{equation}
\label{Equ11delta1}
4U_0\dot{H}\left(1-\delta_1\right)={}-\psi^2+2U_0H^2\delta_1\left(\delta_2+\varepsilon_1-1\right).
\end{equation}

    From Eq.~(\ref{Equ11delta1}), it follows
\begin{equation}
\label{psiviaslr}
\psi^2=2U_0H^2\left[2\varepsilon_1-\delta_1+\delta_1\left(\delta_2-\varepsilon_1\right)\right]
\end{equation}
and
\begin{equation}
\label{chiviaslr}
\chi^2=2U_0\left[2\varepsilon_1-\delta_1+\delta_1\left(\delta_2-\varepsilon_1\right)\right]\,.
\end{equation}

    The scalar spectral index ($n_s$) and tensor spectral index ($n_t$) are linked with the slow-roll parameters as follows~\cite{Guo:2010jr}
\begin{equation}
\label{ns_slr}
n_s=1-2\varepsilon_1-\frac{2\varepsilon_1\varepsilon_2-\delta_1\delta_2}{2\varepsilon_1-\delta_1}=1-2\varepsilon_1-\frac{d\ln(r)}{dN}=1+\frac{d}{dN}\ln\left(\frac{Q}{U_0r}\right),
\end{equation}

\begin{equation}
 n_t = - \frac{2 \varepsilon_1}{1-\varepsilon_1}
 \label{nt}
 \end{equation}
 In some of the EGB model the tensor spectral index is found to be blue tilted~\cite{Oikonomou:2023qfz}. However in our case we expect a red tilted tensor spectral index. From Eq.\ref{nt} it is evident that $n_t$ will always be negative as the slow roll parameter $\varepsilon_1$ is positive during the entire period of inflation. It is rather difficult to achieve the blue tilted $n_t$ in the effective potential construction of the EGB framework.\\
The tensor-to-scalar ratio in terms of slow roll parameters ~\cite{Guo:2010jr},
\begin{equation}
\label{r_slr}
r=8|2\varepsilon_1-\delta_1|.
\end{equation}

    From Eq.~(\ref{chiviaslr}), it follows that
\begin{equation*}
r=\left|\frac{U_0}{\xi_{,\phi}^2Q^2}\delta_1^2-8\delta_1\delta_2+8\delta_1\varepsilon_1\right|\,.
\end{equation*}

 We can find afterwards
\begin{equation}
\label{As_slr}
A_s=\frac{Q}{\pi^2 U_0\, r}\,.
\end{equation}

    On substituting~(\ref{chiviaslr}) into Eq.~(\ref{Equ00delta1}), we get
\begin{equation}
\label{viaslr}
V=U_0H^2\left[6-2\varepsilon_1-5\delta_1-\delta_1\left(\delta_2-\varepsilon_1\right)\right]\,.
\end{equation}

   The exact expression for $\frac{dQ}{dN}$ can be written as:
\begin{equation}
\label{Equ11QNdelta1}
\frac{dQ}{dN}=\frac{Q}{2\,U_0\left(1-\delta_1\right)}\left({}-\chi^2+2\,U_0\delta_1\left(\delta_2+\varepsilon_1-1\right)\right).
\end{equation}


\section{New slow-roll approximations}
\renewcommand{\theequation}{4.\arabic{equation}} \setcounter{equation}{0}
\label{sranew}
To study the stability of de-Sitter solutions of the  model introduced in Eq.( \ref{action1}), an effective potential method has been proposed in ~\cite{Pozdeeva:2019agu} (see also~\cite{Pozdeeva:2020apf,Vernov:2021hxo}):
\begin{equation}
\label{Veff}
V_{eff}(\phi)={}-\frac{U_0^2}{V(\phi)}+\frac{1}{3}\xi(\phi).
\end{equation}

\label{New-slow-roll-approximation}

\subsection{New approximation I}
We obtain the following equation by multiplying Eq.(\ref{Equ00delta1}) with $Q$ and inserting $\psi$ in terms of the slow-roll parameter $\delta_1$:
\begin{equation}
\label{2indel1}
12\,{U_0}\, \left( 1-{\delta_1} \right)Q^2 -2VQ-\frac{\delta_1^2 U_0^2}{4\,\xi^2_{,\phi}}=0\,.
\end{equation}

    Provided, $\delta_1<1$, then Eq.(\ref{2indel1})  will always have a positive root
\begin{equation}
\label{Qdelta1}
Q=\frac{V}{12\,U_0
\, \left(1-\delta_1 \right)}+\frac{\sqrt{{V}^{2}\xi_{,\phi}^{2}+3\,U_0^3\delta_1^2\left(1-\delta_1\right)}}{12\,U_0
\, \left(1-\delta_1 \right)|\xi_{,\phi}|}\,.
\end{equation}

Assuming $\delta_1\ll 1$ and expanding the obtained expression to series with respect to the slow-roll parameter $\delta_1$:
\begin{equation}
\label{Qseries}
Q\approx {\frac{V}{6\,U_0}}+{\frac{V}{6\,U_0}}\delta_1+{\cal{O}}(\delta^2_1)\,.
\end{equation}

    We construct slow-roll approximations with
\begin{equation}
\label{equ00slr}
Q\simeq\frac{V}{6\,U_0}\left(1+\delta_1\right)=\frac{1}{6U_0^2}\left[U_0V+2V\xi_{,\phi}H\psi\right]\,.
\end{equation}

  In order to distinguish the standard slow-roll approximation, we have not neglected $\delta_1$, therefore we should get $\delta_1(\phi)$ to get $Q(\phi)$. To do it we use the following approximation of Eq.~(\ref{Equ11}):
\begin{equation}
\label{equ11slr}
\dot H \simeq{}-\frac{\psi^2+4H^3\xi_{,\phi}\psi}{4U_0\left(1-\delta_1\right)}.
\end{equation}

    Using Eq.~(\ref{delta}) and neglecting the terms that are proportional to $\delta_1^3$, we rewrite~(\ref{equ11slr}) as follows
\begin{equation}
\label{equ11slrdel1}
\dot H\simeq{}-\frac{\delta_1}{4(1-\delta_1)}\left(\frac{U_0\delta_1}{4\xi_{,\phi}^2Q}+2Q\right)\simeq{}-\frac{Q\delta_1}{2}-\frac{U_0\delta_1^2}{16\xi_{,\phi}^2Q}-\frac{Q\delta_1^2}{2}\,.
\end{equation}

  In Eq.~(\ref{Equphi}), terms proportional to $\ddot{\phi}$ and ${\dot{\phi}}^2$ are neglected and Eq.~(\ref{delta}) is used to get the following approximate equation: 
\begin{equation}
\label{equphislr1}
\frac{3U_0\delta_1}{2\xi_{,\phi}}={} -V_{,\phi} -12Q\xi_{,\phi}\left(\dot{H}+Q\right).
\end{equation}

On substituting $Q$ and $\dot H$ from Eqs.~(\ref{equ00slr}) and (\ref{equ11slrdel1}) into Eq.~(\ref{equphislr1}) and neglecting terms proportional to  $\delta_1^n$, where $n\geqslant 2$, we obtain
\begin{equation}
\delta_1(\phi)={}-\frac{2\, V^2\xi_{,\phi}\,{V_{eff}}_{,\phi}}{V^2\xi_{,\phi}^2+3\,U_0^3}\,.
\label{delta1phi}
\end{equation}

Knowing $\delta_1(\phi)$, we get $Q(\phi)$ and $\chi(\phi)$. We get from Eq.~(\ref{equ00slr}) that
\begin{equation}
\label{H2slr}
Q\simeq\frac{V}{6U_0}\left[1-\frac{2V^2\xi_{,\phi}{V_{eff}}_{,\phi}}{V^2\xi_{,\phi}^2+3U_0^3}\right]=\frac{V\left(9U_0^3-6U_0^2\xi_{,\phi}\,V_{,\phi}+\xi_{,\phi}^2V^2\right)}{18U_0\left(3U_0^3+\xi_{,\phi}^2V^2\right)}.
\end{equation}

    Therefore, Eq.~(\ref{delta}) gives
\begin{equation}
\label{apprIequdphidN}
\chi=\frac{U_0\delta_1}{2\xi_{,\phi}Q}\simeq{}-\frac{6U_0^2V{V_{eff}}_{,\phi}}{V^2\xi_{,\phi}^2+3U_0^3-2V^2\xi_{,\phi}{V_{eff}}_{,\phi}}
={}-\frac{6U_0^2\left(3U_0^2V_{,\phi}+\xi_{,\phi}V^2\right)}{V\left(9U_0^3-6U_0^2\xi_{,\phi}V_{,\phi}+\xi_{,\phi}^2V^2\right)}\,,
\end{equation}
hence,
\begin{equation}
\label{apprIdNDdphi}
\frac{dN}{d\phi}= {}-\frac{V^2\xi_{,\phi}^2+3U_0^3-2V^2\xi_{,\phi}{V_{eff}}_{,\phi}}{6U_0^2V{V_{eff}}_{,\phi}}\,.
\end{equation}

 Using Eq.~(\ref{H2slr}), we compute the slow-roll parameters as functions of $\phi$ as:
\begin{equation}
\label{apprIeps1phi}
\varepsilon_1(\phi)={}-\frac{1}{2}\frac{d\phi}{dN}\frac{d\ln(Q)}{d\phi}=\frac{3U_0^2\left(3U_0^2V_{,\phi}+\xi_{,\phi}V^2\right)}{V\left(9U_0^3-6U_0^2\xi_{,\phi}V_{,\phi}+\xi_{,\phi}^2V^2\right)}\,\frac{d\ln(Q)}{d\phi}\\
\end{equation}
where
\begin{equation*}
\frac{d\ln(Q)}{d\phi}=\frac{V_{,\phi}}{V}+\frac{
2\left[\xi_{,\phi}\xi_{,\phi\phi}V^2+\xi_{,\phi}^2VV_{,\phi}-3U_0^2\xi_{,\phi\phi}V_{,\phi}-3U_0^2\xi_{,\phi}V_{,\phi\phi}\right]}
{9U_0^3-6U_0^2\xi_{,\phi}V_{,\phi}+\xi_{,\phi}^2V^2}-\frac{2\xi_{,\phi}\xi_{,\phi\phi}V^2+2{\xi{,\phi}}^2VV_{,\phi}}{3U_0^3+\xi_{,\phi}^2V^2}.
\end{equation*}

We also obtain
\begin{equation}
\label{eps2delta2phi}
\varepsilon_2(\phi)=\frac{U_0\delta_1}{2\xi_{,\phi}Q\varepsilon_1}{\varepsilon_1}_{,\phi}\,,\qquad
\delta_2=\frac{U_0}{2Q\xi_{,\phi}}{\delta_1}_{,\phi}.
\end{equation}

\subsection{New approximation II}
We adopt another method to calculate the $\delta_1(\phi)$ where we use the Eq. (\ref{2indel1}) and neglect the term proportional to $\delta_1^2$ and get a non-zero solution:
\begin{equation}
\label{apprIIH2}
Q=\frac{V}{6U_0(1-\delta_1)}.
\end{equation}

  On using Eq.~(\ref{delta}), we obtain:
\begin{equation}
\label{apprIIdH2dN}
\frac{dQ}{dN}=\frac{V_{,\phi}\,\delta_1}{12\xi_{,\phi}Q(1-\delta_1)}+\frac{V\delta_1\delta_2}{6U_0(1-\delta_1)^2}=\frac{U_0\,V_{,\phi}\,\delta_1}{2\xi_{,\phi}V}+\frac{V\delta_1\delta_2}{6U_0(1-\delta_1)^2}\,.
\end{equation}

    So,
\begin{equation}
\label{epsilon1fromDiff}
\varepsilon_1={}-\frac{3\,U^2_0\,V_{,\phi}}{2 V^2\xi_{,\phi}}\delta_1(1-\delta_1)-\frac{\delta_1\delta_2}{2\,(1-\delta_1)}\,.
\end{equation}

    From Eq.(\ref{delta2}), we obtain
\begin{equation}
\label{del2}
\dot{\psi}\approx\frac{U_0\delta_1}{2\xi_\phi}\left(\delta_2+\varepsilon_1-\frac{3U_0^2\xi_{,\phi\phi}\delta_1}{V\,\xi_{,\phi}^2}\right)\,.
\end{equation}

  We insert formulae given in Eq.( \ref{apprIIH2}), Eq. (\ref{epsilon1fromDiff}), and Eq.( \ref{del2}) into Eq.(\ref{Equphi}), to get
\begin{equation}
\frac{U_0\delta_1}{2\xi_\phi}\left(\delta_2+\varepsilon_1-\frac{3U_0^2\xi_{,\phi\phi}\delta_1}{V\,\xi_{,\phi}^2}\right)+\frac{3U_0\delta_1}{2\xi_{,\phi}}+V_{,\phi}+\frac{12\xi_{,\phi}V^2\,V_{,\phi}}{36\,U_0^2(1-\delta_1)^2}(1-\varepsilon_1)=0
\end{equation}
Now we can construct the linear equation in $\delta_1$ by multiplying the above equation with  $(1-\delta_1)^2$ and assuming the product of two slow-roll parameters are negligible,

\begin{equation}
\frac{9}{2} \left(\frac{U_0^3}{\xi_{,\phi}}-V_{,\phi}\,U_0^2\right) {\delta_1}
+3\,V_{,\phi}\,U_0^2+V^2\xi_{,\phi}=0.
\end{equation}

  Upon solving the above equation, we obtain
\begin{equation}
\label{apprIIdel1phi}
\delta_1(\phi)={}-\frac{2\xi_{,\phi}\left(3U_0^2V_{,\phi}+V^2\xi_{,\phi}\right)}{9U_0^2\left(U_0-\xi_{,\phi}V_{,\phi}\right)}.
\end{equation}

    Expressing $Q$, $\chi$, $N_{,\phi}\,$, and $\varepsilon_1$ in $\phi$ we get: 
\begin{equation}
\label{apprIIH2phi}
Q(\phi)\simeq\frac{V}{6U_0(1-\delta_1(\phi))}=\frac{3{U_0}V(U_0-\xi_{,\phi}V_{,\phi})}{2(9U_0^3-3U_0^2\xi_{,\phi}V_{,\phi}+2\xi_{,\phi}^2V^2)},
\end{equation}
\begin{equation}
\label{apprIIequdphidN}
\chi=\frac{U_0\delta_1}{2\xi_{,\phi}Q}\simeq{}-\frac{2(3U_0^2V_{,\phi}+\xi_{,\phi}V^2)(9U_0^3-3U_0^2\xi_{,\phi}V_{,\phi}+2\xi_{,\phi}^2V^2)}{27U_0^2V{(U_0-\xi_{,\phi}V_{,\phi})}^2},
\end{equation}
\begin{equation}
\label{apprIIequdNdphi}
\frac{dN}{d\phi}={}-\frac{27U_0^2V{\left(U_0-\xi_{,\phi}V_{,\phi}\right)}^2}{2\left(3U_0^2V_{,\phi}+\xi_{,\phi}V^2\right)\left(9U_0^3-3U_0^2\xi_{,\phi}V_{,\phi}+2\xi_{,\phi}^2V^2\right)},
\end{equation}
\begin{equation}
\label{apprIIeps1phi}
\varepsilon_1(\phi)={}-\frac{1}{2}\frac{d\phi}{dN}\frac{d\ln(Q)}{d\phi}=\frac{\left(3U_0^2V_{,\phi}+\xi_{,\phi}V^2\right)\left(9U_0^3-3U_0^2\xi_{,\phi}V_{,\phi}+2\xi_{,\phi}^2V^2\right)}
{27U_0^2V{\left(U_0-\xi_{,\phi}V_{,\phi}\right)}^2}\,\frac{d\ln(Q)}{d\phi}\,,
\end{equation}
where
\begin{equation}
\label{apprIIdH2dphi}
\frac{d\ln(Q)}{d\phi}=\frac{V_{,\phi}}{V}+\frac{\xi_{,\phi\phi}V_{,\phi}+\xi_{,\phi}V_{,\phi\phi}}{\xi_{,\phi}V_{,\phi}-U_0}+\frac{3U_0^2\xi_{,\phi\phi}V_{,\phi}+3U_0^2\xi_{,\phi}V_{,\phi\phi}-4\xi_{,\phi}\xi_{,\phi\phi}V^2
-4\xi_{,\phi}^2VV_{,\phi}}{9U_0^3-3U_0^2\xi_{,\phi}V_{,\phi}+2\xi_{,\phi}^2V^2}\,.
\end{equation}


\section{Inflationary analysis}
\renewcommand{\theequation}{5.\arabic{equation}} \setcounter{equation}{0}
\label{infa}
In the subsequent sections, we will calculate inflationary observables. To study the dynamics of EGB we choose the mutated hilltop inflation model which can be written as:~\cite{Pal:2009sd,Pinhero:2019sbg}
\begin{equation}
    V=V_0 \left[ 1- \text{sech} \left(\alpha \phi  \right) \right]
    \label{mutpot}
\end{equation}

In our analysis we choose the following form of the EGB coupling \cite{Khan:2022odn,Gangopadhyay:2022vgh,Yogesh:2024mpa}
\begin{equation}
    \xi(\phi) = \frac{\xi_1}{V_0} \tanh \left( \xi_2 (\phi) \right)
    \label{xicoupling}
\end{equation}
where $V_0$ is the scale of inflation which can be fixed by matching the scalar power spectrum ($A_s=2.09 \times 10^{-9}$) and $\xi_1$, $\xi_2$ are the coupling constants. Using Eq.(\ref{mutpot}) and Eq.(\ref{xicoupling}) with Eq.(\ref{Veff}), we can write the explicit form of the effective potential:
\begin{equation}
    V_{eff}(\phi)= -\frac{U_0^{2}}{V_0 \left[1-\text{sech} \left( \alpha \phi  \right) \right]}+ \frac{\xi_1 \text{tanh} \left( \xi_2 \phi  \right)}{3 V_0}
    \label{Veff1}
\end{equation}
 In~\cite{Gangopadhyay:2022vgh} the authors have studied the mutated hilltop inflation in standard slow roll approximation in EGB framework as prescribed in~ \cite{Pozdeeva:2020apf,Pozdeeva:2024ihc}. In our analysis, we choose $U_0=\frac{1}{2}$ and $M_P=1$.

\subsection{Numerical Calculation of Inflationary Observables }
\label{numeric_inf_obs}
\begin{figure}[h!]
    \centering
    \includegraphics[width=0.5\linewidth]{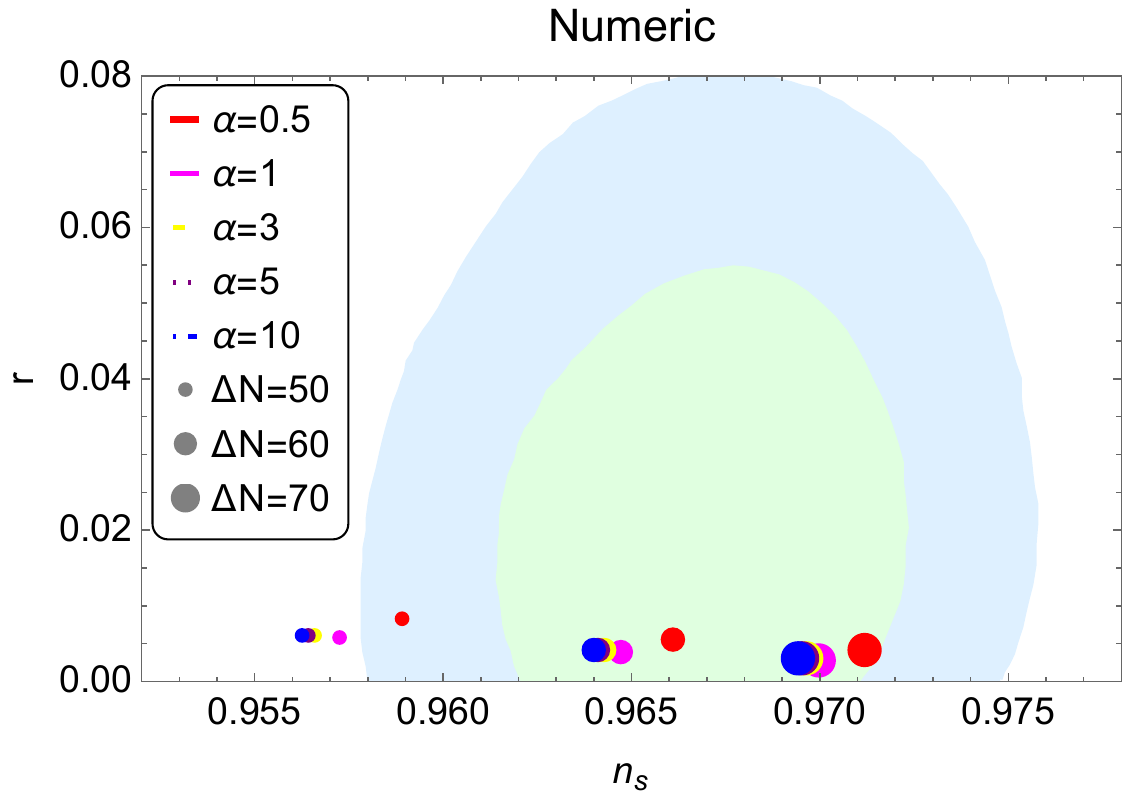}
    \caption{Plot for $r-n_s$ for different values of the potential parameter $\alpha$, using the numerical approach. The light blue shaded region represents the $2-\sigma$ bounds, while the light green shaded region represents the $1-\sigma$ bounds from Planck'18 \cite{Planck:2018jri}. Plot legends are self-explanatory}
    \label{numericnsr}
\end{figure}

Here, we present the numerical analysis of the inflationary observables. We solve the system~ \ref{DynSYSN} numerically to get the solution of $\phi, \chi$ and $Q$ in terms of number of e-folds ($N$). Once we obtained the solution of system~\ref{DynSYSN} it is straightforward to calculate the slow roll parameters using Eq. \ref{epsilon} and \ref{delta}. Substituting the slow roll parameters in Eq.~\ref{ns_slr} and \ref{r_slr} we calculate the scalar spectral index ($n_s$) and tensor to scalar ratio ($r$). It is noteworthy to mention that we obtained the exact values of $n_s$ and $r$ without using any approximations. (see Fig. \ref{numericnsr}). The exact numerical values can be found in Table (\ref{tab_numeric}). We calculate the inflationary observables for different values of potential parameter $\alpha$ and number of e-folds during the inflation from the end of inflation to start of inflation~($\Delta N = N_{end}-N_{start}$).

\begin{table}[h!] \setcounter{table}{0}
\centering
\caption{Table for $r-n_{s}$ obtained by numerically solving Eq.~\ref{DynSYSN}, for different values of potential parameter ($\alpha$) }
\label{tab_numeric}
\renewcommand{\arraystretch}{1.5} 
\setlength{\tabcolsep}{12pt} 
\begin{tabular}{ |c|c|c|c| } 
\hline
$\alpha$ & $\Delta N$ & $n_{s}$ & $r$ \\
\hline
    & 50 & 0.958910 & 0.00815633 \\
 0.5   & 60 & 0.966109 & 0.00564497 \\
    & 70 & 0.971168 & 0.00414538 \\
\hline
\hline
   & 50 & 0.957258 & 0.0057446 \\
   1 & 60 & 0.964772 & 0.00390001 \\
    & 70 & 0.969969 & 0.00281791 \\
\hline
\hline
    & 50 & 0.956613 & 0.0060394 \\
   3 & 60 & 0.964257 & 0.0040774 \\
    & 70 & 0.969625 & 0.00293448 \\
\hline
\hline
    & 50 & 0.956410 & 0.00609666 \\
   5 & 60 & 0.964120 & 0.00410897 \\
    & 70 & 0.969526 & 0.00295379 \\
\hline
\hline
    & 50 & 0.956235 & 0.00614665 \\
  10  & 60 & 0.964001 & 0.00413673 \\
    & 70 & 0.969439 & 0.00297072 \\
\hline
\end{tabular}
\end{table}

\begin{figure}[h!]
    \centering
    \includegraphics[width=0.5\linewidth]{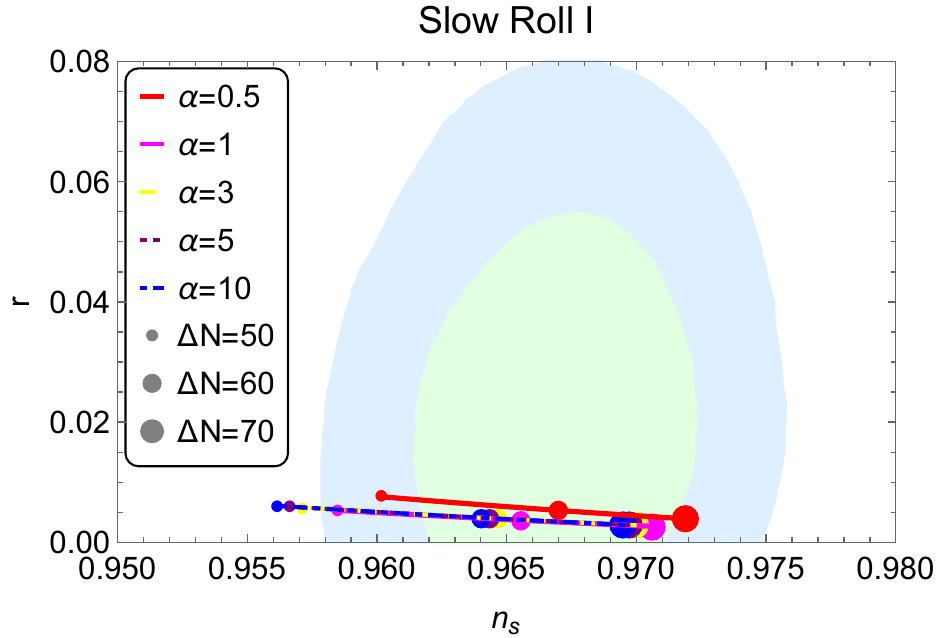}
    \caption{Plot for $r-n_s$ for different values of the potential parameter $\alpha$, using the Slow-Roll Approximation -I. The light blue shaded region represents the $2-\sigma$ bounds, while the light green shaded region represents the $1-\sigma$ bounds from Planck'18 \cite{Planck:2018jri}. Plot legends are self-explanatory}
    \label{rnsnslow1}
\end{figure}

\subsection{Inflationary Observables in Slow-Roll I}
In this section, we present the inflationary analysis of Mutated Hilltop inflation models considering the Slow-Roll Approximation-I. Employing formulae given in Eq.(\ref{ns_slr})--(\ref{As_slr}), we can obtain $n_s(\phi)$, $r(\phi)$, and $A_s(\phi)$ by knowing the slow-roll parameters as functions of $\phi$.
The aforementioned methods may now be applied to the particular instance of the mutated hilltop model. Using the new approximation-I as provided in \ref{App1}, one may obtain the analytical equations for the slow-roll parameters as a function of $\phi$.
\begin{figure}[h!]{\includegraphics[width=0.5\textwidth]{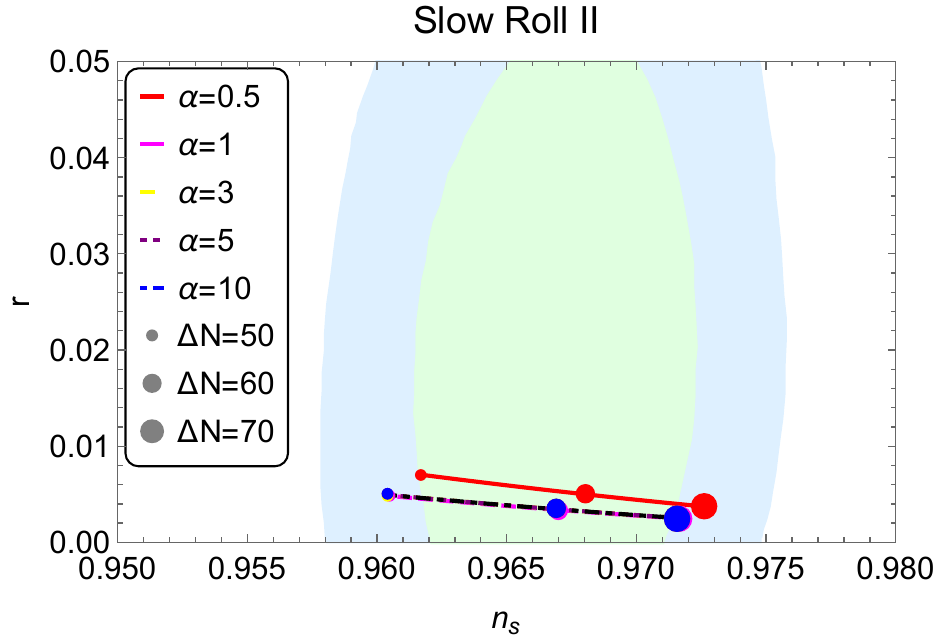}}
\caption{Plot for $r-n_s$ for different values of the potential parameter $\alpha$, using the Slow-Roll Approximation -II. The light blue shaded region represents the $2-\sigma$ bounds, while the light green shaded region represents the $1-\sigma$ bounds on $n_s$ from Planck'18 \cite{Planck:2018jri}. Plot legends are self-explanatory.} 
\label{rnsslow2}
\end{figure}


The results may appear cumbersome, but it's important to recognize that these techniques still allow for the derivation of slow roll parameters analytically. 
 First, we need to calculate the slow roll parameters $\varepsilon_1$ and $\delta_1$ from Eq.(\ref{apprIeps1phi} ) and Eq.(\ref{delta1phi}), after that, it is straightforward to calculate other slow roll parameters ($\varepsilon_2,\delta_2$) using Eq.(\ref{eps2delta2phi}). When studying inflation a crucial quantity to be taken care of is the number of e-folds $N$. To calculate the number of e-folds during inflation ($N$) we solve Eq.(\ref{apprIdNDdphi}) numerically. The initial field value to solve Eq. (\ref{apprIdNDdphi}) can be obtained by imposing the end of inflation condition ($\varepsilon_1=1$). We adopt the same prescription as mentioned in \cite{Pozdeeva:2024ihc}, where we take $N=60$ as the end of inflation and the beginning of inflation is at $N=0$.  
 Equipped with all the necessary equations one can compute the spectral index ($n_S$), tensor to the scalar ratio ($r$), and the amplitude of the scalar power spectrum ($A_s$) using Eq.(\ref{ns_slr}), Eq.(\ref{r_slr}), and Eq.(\ref{As_slr} )respectively. In our analysis we take $\xi_1=5$ and $\xi_2 =0.4$, these values of the coupling parameters are chosen so that the constraints on $n_s$ and $r$ do not contradict the recent CMB observations. We calculate the inflationary observables for different values of $\Delta N$, here $\Delta N$ signifies the number of e-folds during the inflation from the end of inflation to start of inflation~($\Delta N = N_{end}-N_{start}$).  Whereas, $V_0$ can be calculated by fixing $A_s=2.09\times10^{-9}$ at the pivot scale.  Whereas compatibility of our models with the current observations can be found in Fig. \ref{rnsnslow1}, it is evident that in the EGB background, all the inflationary observables are well inside the {\it Planck'18} bounds~\cite{Planck:2018jri}. Whereas the exact numerical values of $r, n_s$ with different values of $\alpha$ and $\Delta N$ can be found in Table \ref{tabslow1}


\begin{table}[h]
\centering
\begingroup
\setlength{\tabcolsep}{10pt} 
\renewcommand{\arraystretch}{1.5} 

\begin{minipage}{0.45\textwidth}
\centering
\caption{Table for $r-n_{s}$ for different values of potential parameters in Slow-Roll I}
\label{tabslow1}
\begin{tabular}{ |c|c|c|c| } 
\hline
$\alpha$ & $\Delta N$ & $n_{s}$ & $r$ \\
\hline
    & 50 & 0.960158 & 0.0075926 \\
 0.5   & 60 & 0.967003 & 0.00530987 \\
    & 70 & 0.971851 & 0.0039292 \\
\hline
\hline
   & 50 & 0.958461 & 0.00534847 \\
   1 & 60 & 0.965578 & 0.00366793 \\
    & 70 & 0.970618 & 0.00267038 \\
\hline
\hline
    & 50 & 0.957098 & 0.0057767 \\
   3 & 60 & 0.964635 & 0.00392139 \\
    & 70 & 0.969925 & 0.00284906 \\
\hline
\hline
    & 50 & 0.956617 & 0.0059521 \\
   5 & 60 & 0.964308 & 0.00401921 \\
    & 70 & 0.969689 & 0.00289418 \\
\hline
\hline
    & 50 & 0.956181 & 0.00607325 \\
  10  & 60 & 0.964013 & 0.00408634 \\
    & 70 & 0.969476 & 0.00293517 \\
\hline
\end{tabular}
\end{minipage}%
\hspace{0.05\textwidth}
\begin{minipage}{0.45\textwidth}
\centering
\caption{Table for $r-n_{s}$ for different values of potential parameters in Slow-Roll II}
\label{tabslow2}

\begin{tabular}{ |c|c|c|c| } 
\hline
$\alpha$ & $\Delta N$ & $n_{s}$ & $r$ \\
\hline
 & 50 & 0.96165 & 0.00704954 \\
  0.5  & 60 & 0.968044 & 0.00499055 \\
    & 70 & 0.972619 & 0.00372546 \\
\hline
\hline
 & 50 & 0.96052 & 0.00481734 \\
   1 & 60 & 0.967017 & 0.00336088 \\
    & 70 & 0.97168 & 0.00247721 \\
\hline
\hline
 & 50 & 0.960383 & 0.00494442 \\
  3  & 60 & 0.966913 & 0.00344483 \\
    & 70 & 0.971596 & 0.00253641 \\
\hline
\hline
 & 50 & 0.960373 & 0.00494687 \\
   5 & 60 & 0.966906 & 0.00344625 \\
    & 70 & 0.971591 & 0.00253731 \\
\hline
\hline
  & 50 & 0.960373 & 0.00494692 \\
   10 & 60 & 0.966906 & 0.00344627 \\
    & 70 & 0.971591 & 0.00253732 \\
\hline
\end{tabular}
\end{minipage}

\endgroup
\end{table}

\subsection{Inflationary Observables in Slow-Roll II}

In this section, we discuss the Mutated Hilltop inflation models within the framework of Slow Roll Approximation II. We employ the same method for calculating inflationary observables as in Slow Roll Approximation I. First, we calculate the slow roll parameters ($\varepsilon_1, \varepsilon_2, \delta_1, \delta_2$). Using equations (\ref{apprIIeps1phi} )and (\ref{apprIIdel1phi}), we compute $\varepsilon_1$ and $\delta_1$, followed by the straightforward determination of $\varepsilon_2$ and $\delta_2$ using Eq.(\ref{eps2delta2phi})~(see~\ref{app2}) for explicit expression of slow roll parameters). We then calculate the number of e-folds, $N$, using equation (\ref{apprIIequdNdphi}), applying the same technique as in the previous subsection (Slow Roll Approximation I).

With these preliminaries in place, we proceed to compute the scalar spectral index ($n_s$), tensor-to-scalar ratio ($r$), and amplitude of the scalar power spectrum ($A_s$) using equations (\ref{ns_slr}), (\ref{r_slr}), and (\ref{As_slr}), respectively. As before, we set $\xi_1 =5$ and $\xi_2 = 0.4$, choosing these coupling parameters to ensure that the resulting constraints on $n_s$ and $r$ are consistent with current CMB bounds. Similar to slow-roll approximation I we calculate the inflationary observables for different values of $\Delta N$, here $\Delta N$ signifies the number of e-folds during the inflation from the end of inflation to start of inflation~($\Delta N = N_{end}-N_{start}$).  The value of $V_0$ is then fixed by matching $A_s = 2.09 \times 10^{-9}$ at the pivot point. Compatibility of our models with the current observations can be found in Fig. \ref{rnsslow2}, it is evident that in the EGB background, all the inflationary observables are well inside the {\it Planck'18} bounds~\cite{Planck:2018jri}. Exact numerical values of $r, n_s$ with different choices of $\alpha$ and $\Delta N$ are presented in Table \ref{tabslow2}

\section{Reheating Parameters}
\renewcommand{\theequation}{6.\arabic{equation}} \setcounter{equation}{0}
\label{sec:reh}
Succeeding the inflationary epoch, the universe enters a super-cooled state. In the standard cold inflationary scenario, the only surviving degree of freedom is the inflaton field itself. However, to repopulate the universe, a reheating phase is necessary \cite{PhysRevLett.48.1437,Kofman:1994rk,Martin:2014nya,Martin:2014nya}. During this phase, the inflaton field transfers its energy to other degrees of freedom, transforming the universe from a super-cooled state into a hot thermal bath of relativistic particles. The concept of reheating was first introduced in \cite{Kofman:1994rk}.

In the standard cold inflationary scenario, several reheating mechanisms have been proposed in the literature. One such method is perturbative decay, where the inflaton field reaches the bottom of its potential and decays into other elementary particles \cite{Kofman:1997yn, Shtanov:1994ce, Bassett:2005xm, Rehagen:2015zma}. These particles interact and eventually reach equilibrium at a temperature known as the reheating temperature ($T_{re}$). More advanced mechanisms include non-perturbative methods like parametric resonance, tachyonic instability \cite{Felder:2001kt}, and preheating \cite{Lozanov:2019jxc, Kofman:1997pt}. The initial stage of reheating is often associated with preheating, a phase more efficient than perturbative reheating due to its exponential particle production.

However, it has been shown that the reheating era can be analyzed without delving into its detailed dynamics \cite{Cook:2015vqa}. Since there is no direct observational constraint on the reheating temperature, an indirect analysis of this phase can be highly beneficial. This approach allows us to estimate the thermalization temperature using inflationary observables, providing a new method to constrain different inflationary models. Besides the reheating temperature ($T_{re}$) and the equation of state parameter ($\omega_{re}$), another key quantity is the duration of reheating ($N_{re}$), which measures the expansion of the universe from the end of inflation to the end of the reheating era \cite{Cook:2015vqa,Khan:2022odn,Gangopadhyay:2022vgh,Adhikari:2019uaw,Mishra:2021wkm,Gialamas:2024jeb,Gialamas:2019nly,Yogesh:2024vcl}.

Adopting, $w_{re}$ to be constant during the reheating period, the energy density of the universe at the end of inflation can be expressed as a function of the scale factor through $\rho \propto a^{-3(1+w)}$, formulated as follows:
\begin{equation}
    \frac{\rho_{end}}{\rho_{re}} = \left(\frac{a_{end}}{a_{re}} \right)^{-3(1+w_{re})},
    \label{re1}
\end{equation}
The subscript $end$ is for the end of inflation, while $re$ denotes the end of the reheating epoch. Utilizing the end of inflation condition ($\varepsilon=1$), we deduce $\rho_{end}= (3/2) V_{end}$ and from Eq. (\ref{re1}), we compute
\begin{equation}
    N_{re} = \frac{1}{3(1+w_{re})} \ln \left(\frac{\rho_{end}}{\rho_{re}} \right)= \frac{1}{3(1+w_{re})} \ln \left(\frac{3}{2}\frac{V_{end}}{\rho_{re}} \right),
    \label{re2}
\end{equation}
Also, we know:
\begin{equation}
\rho_{re} = \frac{\pi^2}{30} g_{re} T_{re}^4.
\label{re3}
\end{equation}
Where $g_{re}$ is for the number of relativistic degrees of freedom at the termination of the reheating phase.

Using Eq.(\ref{re2}) and Eq.(\ref{re3}) and following  \cite{Cook:2015vqa, Cai:2015soa, Gong:2015qha}, we can express $T_{re}$ and $N_{re}$ as :
\begin{equation}
N_{re} = \frac{1}{3(1+w_{re})} \ln \left(\frac{30 \cdot \frac{3}{2}  V_{end}}{\pi^2 g_{re} T_{re}^4 } \right)
\label{re4}
\end{equation}
Accepting that entropy remains conserved from the reheating era to today, we write  
\begin{equation}
T_{re}= T_0 \left(\frac{a_0}{a_{re}} \right) \left(\frac{43}{11 g_{re}} \right)^{\frac{1}{3}}=T_0 \left(\frac{a_0}{a_{eq}} \right) e^{N_{RD}} \left(\frac{43}{11 g_{re}} \right)^{\frac{1}{3}},
\label{re5}
\end{equation}
with $N_{RD}$ being the number of e-folds during radiation era, since $e^{-N_{RD}}\equiv a_{re}/a_{eq}$. The ratio $a_{0}/a_{eq}$ can be expressed as 
\begin{equation}
\frac{a_0}{a_{eq}} = \frac{a_0 H_{k}}{k} e^{-N_{k}} e^{- N_{re}} e^{- N_{RD}}\
\label{re6}
\end{equation}

We know for the modes at the horizon exit we can write, $k_{}=a_{k} H_{k}$ and using the Eq.~(\ref{re4}), Eq.(\ref{re5}) and Eq. (\ref{re6}), assuming $w_{re} \neq \frac{1}{3}$ and $g_{re} \approx 226$ (degrees of freedom in a supersymmetric model), we can derive the expression for $N_{re}$
\begin{equation}
N_{re}= \frac{4}{ (1-3w_{re} )}   \left[61.488  - \ln \left(\frac{ V_{end}^{\frac{1}{4}}}{ H_{k} } \right)  - N_{k}   \right]
\label{re7}
\end{equation}
Using Planck's pivot  ($k$) of order $0.05 \; \mbox{Mpc}^{-1}$ we find $T_{re}$:
\begin{equation}
T_{re}= \left[ \left(\frac{43}{11 g_{re}} \right)^{\frac{1}{3}}    \frac{a_0 T_0}{k_{}} H_{k} e^{- N_{k}} \left[\frac{3^2 \cdot 5 V_{end}}{\pi^2 g_{re}} \right]^{- \frac{1}{3(1 + w_{re})}}  \right]^{\frac{3(1+ w_{re})}{3 w_{re} -1}}.
\label{re8}
\end{equation}
The calculations of $H_{k}$, $N_{k}$ (e-folds throughout the inflation), and $V_{end}$ (value of the potential at the end of inflation) are necessary to assess $N_{re}$ and $T_{re}$, as can be seen from Eq. (\ref{re7}) and Eq. (\ref{re8}). With the scalar power spectrum and the tensor-to-scalar ratio, we may write
\begin{equation}
{H_k}=\sqrt{\frac{1}{2} \pi ^2 A_{s}  r}.
\label{Hk}
\end{equation}

It is evident from Eq.(\ref{Hk}) that $H_k$ is a function of the scalar ratio ($r$) to tensor. The CMB measurement informs us that the tensor-to-scalar ratio has only an upper bound. To accurately examine the reheating dynamics, we must represent $H_k$ in terms of the spectral index ($n_s$). The relationship between $r,n_s$ and $N_k$\cite{Adhikari:2019uaw,Khan:2022odn} may be used to do this. As noted in~\cite{Cook:2015vqa} reheating dynamics is not very sensitive to the change in  $A_s$ with $n_s$ we keep $A_s(k_0)= 2.0989\times10^{-9}$. 

\begin{figure}[!htb]
\centering
\includegraphics[width=8cm,height=8cm]{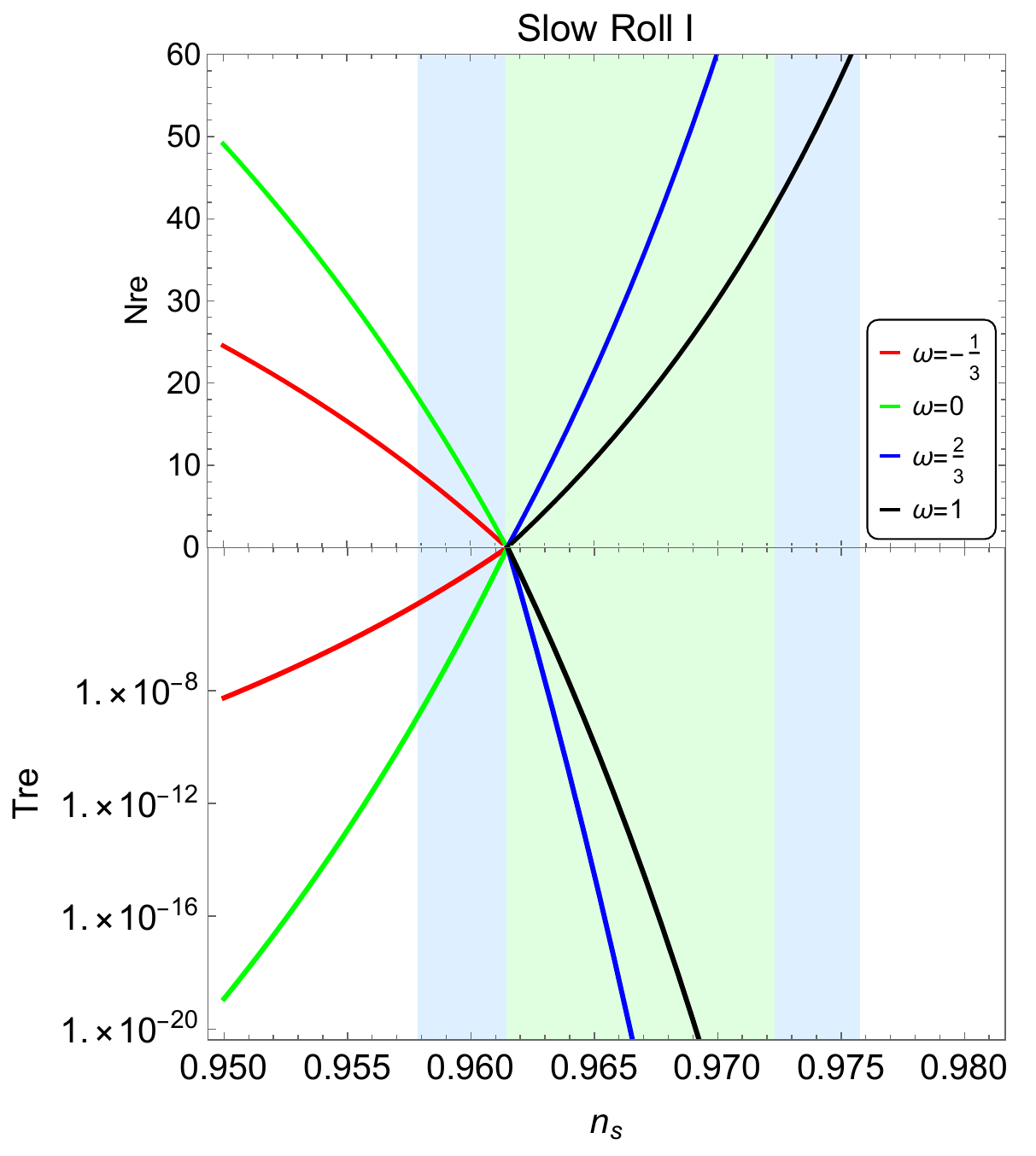}
\includegraphics[width=8cm,height=8cm]{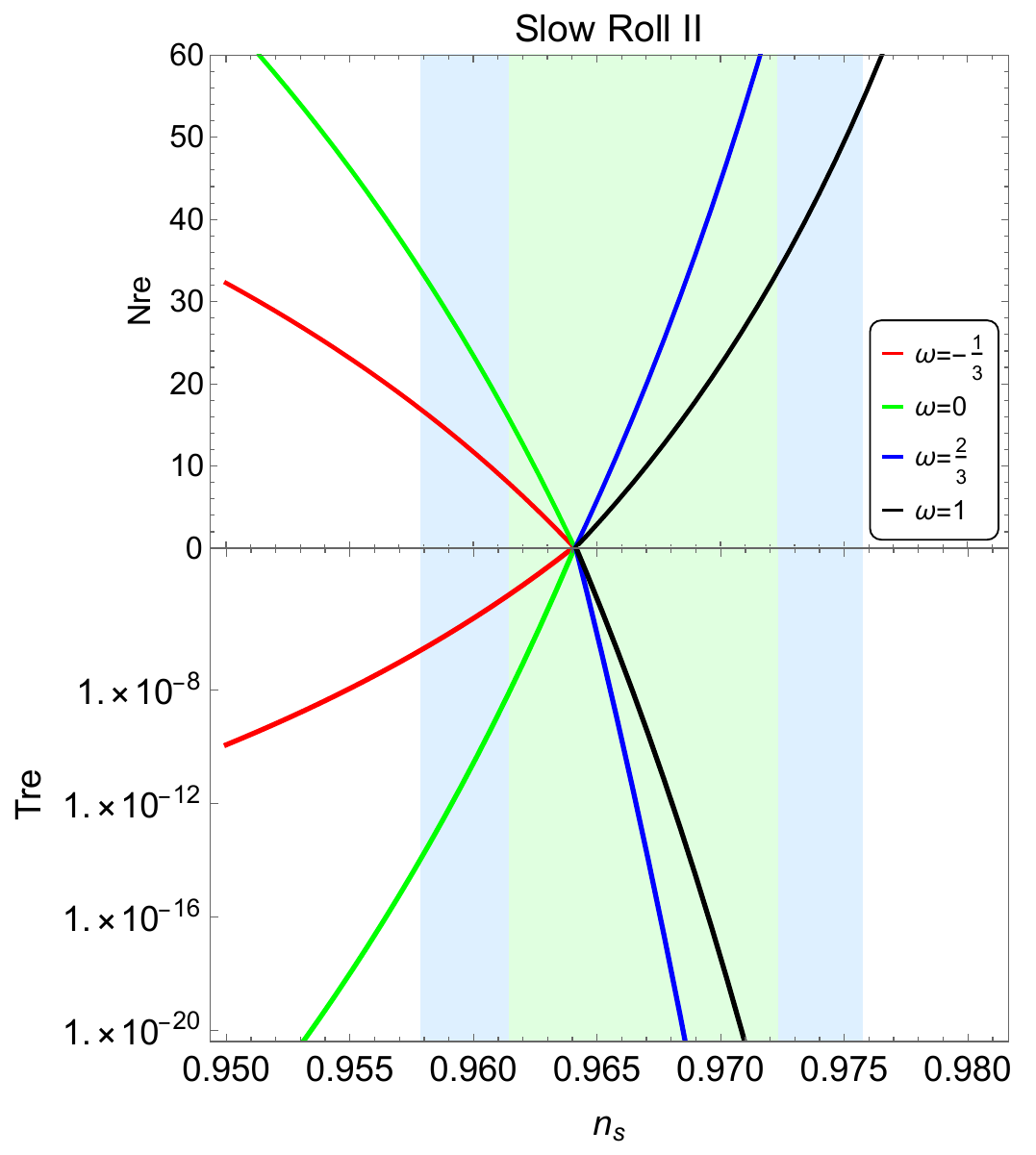}
\caption{Plots illustrating $N_{re}$ and $T_{re}$ for different choices of $\omega_{re}$ are presented. The light green shaded area shows the $1-\sigma$ bounds on $n_s$ from Planck'18 \cite{Planck:2018jri}, while the light blue shaded area is for the $2-\sigma$ bounds. Lastly, the inset of the graphs includes the self-explanatory color coding for $\omega_{re}$.
\textbf{Left Panel:} The results for slow-roll approximation $\text{I}$ is shown, while in the  
\textbf{Right Panel:}  The results for slow-roll approximation $\text{II}$ is shown.}
\label{reheating_plots}
\end{figure}

 The variation of $N_{re}$ and $T_{re}$ with $n_s$ for mutated hilltop inflation is demonstrated in Fig.~\ref{reheating_plots}. Instantaneous reheating is shown by the convergence points on the $T_{re}$ and $N_{re}$ plots, where $N_{re}=0$ in fig.~\ref{reheating_plots}.

\section{Conclusions and Discussions}
\label{discussions}
The theoretically well-motivated models can be tested against future observations such as {\it CMB-S4} ~\cite{CMB-S4:2016ple} and {\it COrE} ~\cite{CORE:2016ymi}, which could measure the spectral tilt very precisely ($\Delta n_s \sim 0.002$) and provide constraints on the primordial tensor modes. Furthermore, the conceptual advancements in the field of modified gravity during the past several years have created an opportunity to examine several fascinating aspects. Novel advancements in EGB gravity, such as the slow-roll approximation techniques presented in ~\cite{Pozdeeva:2024ihc}, demand a more thorough examination. In this paper, we study the Mutated Hilltop Inflation in the new slow roll approximations.  We show that for a wide range of potential parameters $\alpha=(0.5,1,3,5,10)$ along the suitable choice of EGB coupling the constraints imposed tensor to a scalar ratio ($r$) and spectral index ($n_s$) by CMB observations ~\cite{Planck:2018jri} can be easily satisfied. In slow-roll approximation -I (see figs. \ref{rnsnslow1})  it is clear that for $\Delta N=(60,70)$ all the values of $r$ and $n_s$ for different values of $a\alpha$ are inside the $2-\sigma$ bound imposed by CMB. Whereas for $\Delta N=50$ the results are outside the $2-\sigma$ region for certain values of $\alpha$~(see Table~\ref{tabslow1}). In slow-roll approximation II (see figs. \ref{rnsslow2}) it is clear that for $\Delta N=(50,60,70)$ all the values of $r$ and $n_s$ for different values of $\alpha$ are inside the $2-\sigma$ bound imposed by CMB. Whereas, the inflationary observables that are obtained by numerically solving the system~\ref{DynSYSN} are given in fig. \ref{numericnsr} and Table ~\ref{tab_numeric}. For $\Delta N=(60,70)$ the numerical solution obtained for different values of $\alpha$ are well inside the $2-\sigma$ bound imposed by CMB. However, for $\Delta N=50$ and for some values of $\alpha$, the results are outside $2-\sigma$ which is also the case with slow roll approximation I. Interestingly, the inflationary observables obtained numerically are closer to the results obtained in slow-roll approximation I.  Nevertheless, the study of the inflationary epoch is incomplete at least in the standard cold inflationary scenario, thus in this manuscript, we have also studied the reheating dynamics.  We have studied the reheating for four different equations of state parameters~($\omega_{re}$). It is clear from the  Fig.~\ref{reheating_plots} that $\omega_{re}=(2/3,1)$ are well inside the $2-\sigma$ bound on $n_S$ for entire range to $T_{re}$ from instantaneous reheating temperature to the BBN temperature($\approx 1 \text{Mev}$). As we mentioned, the results for $n_s$ and $r$ obtained through numerical analysis are similar to those obtained in the slow-roll approximation I. Therefore, we expect the reheating behavior in the numerical calculation to be similar to that in the slow-roll approximation I.
We had three motives for studying the Mutated Hilltop Potential, which are summarized below.  
\begin{itemize}
    \item Demonstrating that it is possible to effectively rescue a theoretically well-motivated model utilizing a non-standard gravitational backdrop of EGB gravity.
    \item Apply the recently suggested concept of slow-roll approximation to the mutated hilltop model and evaluate the variation in predictability based on the chosen approximation technique.
    \item Ultimately, considering a more practical scenario of widespread reheating, the ability of inflation to make accurate predictions sheds light on a model inside the Einstein-Gauss-Bonnet (EGB) framework. 
\end{itemize}

We want to carry out a few additional investigations in the EGB inflationary framework to see how Warm inflation dynamics are included \cite{Bastero-Gil:2016qru,Bastero-Gil:2017wwl}. Utilizing this new prescription might be an extremely intriguing avenue to follow: "Production of Primordial Black Holes and Gravitational Waves in case of Warm Inflationary Dynamics" \cite{Gangopadhyay:2020bxn, Basak:2021cgk, Correa:2023whf,Correa:2022ngq} in the EGB backdrop. We will come back to these issues in the near future.

\section*{Acknowledgments} 
The authors would like to thank Mayukh Raj Gangopadhyay and Abolhassan Mohammadi for the useful discussion.

\bibliographystyle{apsrev4-1}
\bibliography{MH_EGB.bib}


\newpage
\appendix

\section{Explicit Analytic Forms of Slow-Roll Parameters}
\label{App}
\subsection{\underline{New Slow Roll Approximation I}}
\label{App1}
The slow-roll parameters in the case of slow-roll approximation I are given below:
\begin{align}
    \delta_1=-\frac{\xi _1 \xi _2 (\text{sech}(\alpha  \phi )-1)^2 \text{sech}^2\left(\xi _2 \phi \right) \left(3 \alpha  \sinh (\alpha  \phi ) \text{csch}^4\left(\frac{\alpha  \phi }{2}\right)+16 \xi _1 \xi _2 \text{sech}^2\left(\xi _2 \phi \right)\right)}{96 \xi _1^2 \xi _2^2 \sinh ^4\left(\frac{\alpha  \phi }{2}\right) \text{sech}^2(\alpha  \phi ) \text{sech}^4\left(\xi _2 \phi \right)+9}
\end{align}
\begin{align}
    \varepsilon_1 =& \Bigg(  3 \text{sech}^2(\alpha  \phi ) \Big(16 \xi _1 \xi _2 \sinh ^4\left(\frac{\alpha  \phi }{2}\right) \text{sech}(\alpha  \phi ) \text{sech}^2\left(\xi _2 \phi \right)+3 \alpha  \tanh (\alpha  \phi )\Big) \times \textbf{D1} \Bigg)\Bigg/ \nonumber \\ & \quad \Bigg( 2 (\text{sech}(\alpha  \phi )-1)^2 \left(32 \xi _1^2 \xi _2^2 \sinh ^4\left(\frac{\alpha  \phi }{2}\right) \text{sech}^2(\alpha  \phi ) \text{sech}^4\left(\xi _2 \phi \right)+3\right)  \Bigg(4 \xi _1 \xi _2 \text{sech}(\alpha  \phi ) \text{sech}^2\left(\xi _2 \phi \right)+9  8 \xi _1 \xi _2 \sinh ^4\left(\frac{\alpha  \phi }{2}\right) \nonumber \\ & \quad\text{sech}(\alpha  \phi ) \text{sech}^2\left(\xi _2 \phi \right)-3 \alpha  \tanh (\alpha  \phi )  \Bigg)^2   \Bigg)
\end{align}

\begin{align}
    \delta_2 =& \Bigg( 9 \text{csch}^4\left(\frac{\alpha  \phi }{2}\right) \text{sech}^2(\alpha  \phi ) \cosh ^2\left(\xi _2 \phi \right) \bigg( 6 \alpha ^2 (\cosh (2 \alpha  \phi )-3) \text{csch}^6\left(\frac{\alpha  \phi }{2}\right) \cosh ^6\left(\xi _2 \phi \right)  \nonumber \\ & \quad + 4 \alpha  \xi _2 \coth \left(\frac{\alpha  \phi }{2}\right) \text{csch}^4\left(\frac{\alpha  \phi }{2}\right) \cosh ^4\left(\xi _2 \phi \right) \left(6 \cosh (\alpha  \phi ) \sinh \left(2 \xi _2 \phi \right)-16 \xi _1 (\cosh (\alpha  \phi )-1)\right)  \nonumber \\ & \quad + 32 \xi _1 \xi _2^2 \text{csch}^2\left(\frac{\alpha  \phi }{2}\right) \text{sech}(\alpha  \phi ) \cosh ^2\left(\xi _2 \phi \right) \left(4 \alpha ^2 \xi _1 (\cosh (\alpha  \phi )+2)-\sinh \left(2 \phi  \left(\alpha -\xi _2\right)\right)+\sinh \left(2 \phi  \left(\alpha +\xi _2\right)\right)+2 \sinh \left(2 \xi _2 \phi \right)\right) \nonumber \\ & \quad -  256 \alpha  \xi _1^2 \xi _2^3 \coth \left(\frac{\alpha  \phi }{2}\right) \text{sech}(\alpha  \phi ) \sinh \left(2 \xi _2 \phi \right) \bigg) \Bigg)\Bigg/ \Bigg( 8 \Big( 32 \xi _1^2 \xi _2^2 \text{sech}^2(\alpha  \phi )+3 \text{csch}^4\left(\frac{\alpha  \phi }{2}\right) \cosh ^4\left(\xi _2 \phi \right) \Big) \times \nonumber \\ & \quad \bigg( 32 \xi _1^2 \xi _2^2 \text{sech}^2(\alpha  \phi )+9 \text{csch}^4\left(\frac{\alpha  \phi }{2}\right) \cosh ^4\left(\xi _2 \phi \right)-12 \alpha  \xi _1 \xi _2 \tanh (\alpha  \phi ) \text{csch}^4\left(\frac{\alpha  \phi }{2}\right) \text{sech}(\alpha  \phi ) \cosh ^2\left(\xi _2 \phi \right)   \bigg)  \Bigg) 
\end{align}

\begin{align}
\varepsilon_2 =& \Bigg[ 6 \Big( 3 \alpha  \sinh (\alpha  \phi ) \text{csch}^4\left(\frac{\alpha  \phi }{2}\right)+16 \xi _1 \xi _2 \text{sech}^2\left(\xi _2 \phi \right)\Big) \Big( \xi _1^2 \xi _2^2 (\text{sech}(\alpha  \phi )-1)^2 \text{sech}^4\left(\xi _2 \phi \right)+\frac{3}{8} \Big) \times \nonumber \\ & \quad \Big( 32 \xi _1^2 \xi _2^2 \sinh ^4\left(\frac{\alpha  \phi }{2}\right) \text{sech}^2(\alpha  \phi ) \text{sech}^4\left(\xi _2 \phi \right)-12 \alpha  \xi _1 \xi _2 \tanh (\alpha  \phi ) \text{sech}(\alpha  \phi ) \text{sech}^2\left(\xi _2 \phi \right)+9 \Big) \times \nonumber \\ & \quad \Bigg( \left( \text{sech}(\alpha  \phi )-1  \right) \text{sech}^2(\alpha  \phi ) \times \textbf{A1}\times \textbf{B1} \times \textbf{C1} \Bigg(27 \alpha ^2+1536 \xi _1^2 \xi _2^3 \sinh ^5\left(\frac{\alpha  \phi }{2}\right) \text{sech}^4\left(\xi _2 \phi \right) \times \nonumber \\ & \quad \bigg(- \Big(  \xi _2 \sinh \left(\frac{\alpha  \phi }{2}\right) \left(2 \cosh \left(2 \xi _2 \phi \right)-3\right) \text{sech}^2\left(\xi _2 \phi \right) \Big) + \frac{1}{2} \alpha \left(5 \cosh \left(\frac{\alpha  \phi }{2}\right)+\cosh \left(\frac{3 \alpha  \phi }{2}\right)\right) \text{sech}(\alpha  \phi ) \tanh \left(\xi _2 \phi \right) \bigg) \times \nonumber \\ & \quad +   1024 \alpha  \xi _1^4 \xi _2^4 \sinh ^8\left(\frac{\alpha  \phi }{2}\right) \text{sech}^3(\alpha  \phi ) \text{sech}^8\left(\xi _2 \phi \right) \Big(\alpha  (5 \text{sech}(\alpha  \phi )+4)-8 \xi _2 \sinh (\alpha  \phi ) \tanh \left(\xi _2 \phi \right)\Big) + \nonumber \\ & \quad 36 \alpha  \xi _1 \xi _2 \sinh \left(\frac{\alpha  \phi }{2}\right) \text{sech}^2\left(\xi _2 \phi \right) \Bigg(\alpha ^2 \left(-12 \cosh \left(\frac{\alpha  \phi }{2}\right)+5 \cosh \left(\frac{3 \alpha  \phi }{2}\right)+\cosh \left(\frac{5 \alpha  \phi }{2}\right)\right) \text{sech}^2(\alpha  \phi ) \nonumber \\ & \quad - 8 \xi _2^2 \sinh ^2\left(\frac{\alpha  \phi }{2}\right) \cosh \left(\frac{\alpha  \phi }{2}\right) \left(\cosh \left(2 \xi _2 \phi \right)-2\right) \text{sech}^2\left(\xi _2 \phi \right)+ \alpha  \text{sech}(\alpha  \phi ) \xi _2 \tanh \left(\xi _2 \phi \right) \times \nonumber \\ & \quad \Big( 14 \sinh \left(\frac{\alpha  \phi }{2}\right)+5 \sinh \left(\frac{3 \alpha  \phi }{2}\right)-\sinh \left(\frac{5 \alpha  \phi }{2}\right)  \Big)\Bigg) + 384 \alpha  \xi _1^3 \xi _2^3 \sinh ^5\left(\frac{\alpha  \phi }{2}\right) \text{sech}^2(\alpha  \phi ) \text{sech}^6\left(\xi _2 \phi \right) \times \nonumber \\ & \quad \Big( \alpha ^2 \left(-2 \cosh \left(\frac{\alpha  \phi }{2}\right)+7 \cosh \left(\frac{3 \alpha  \phi }{2}\right)+\cosh \left(\frac{5 \alpha  \phi }{2}\right)\right) \text{sech}^2(\alpha  \phi )+ 8 \cosh \left(\frac{\alpha  \phi }{2}\right) \left(3 \cosh \left(2 \xi _2 \phi \right)-4\right) \times \nonumber \\ & \quad \xi _2^2 \sinh ^2\left(\frac{\alpha  \phi }{2}\right) \text{sech}^2\left(\xi _2 \phi \right) \left.\alpha  \xi _2 \left(2 \sinh \left(\frac{\alpha  \phi }{2}\right)-3 \left(3 \sinh \left(\frac{3 \alpha  \phi }{2}\right)+\sinh \left(\frac{5 \alpha  \phi }{2}\right)\right)\right) \text{sech}(\alpha  \phi ) \tanh \left(\xi _2 \phi \right)\right)  \Big)                     
\Bigg) \nonumber \\ & \quad - 2 \alpha  (\text{sech}(\alpha  \phi )-1) \tanh (\alpha  \phi ) \times \textbf{A1}\times \textbf{B1} \times \textbf{C1} \times \textbf{D1} + 2 \alpha  \text{sech}(\alpha  \phi ) \tanh (\alpha  \phi )  \times \textbf{A1}\times \textbf{B1} \times \textbf{C1} \times \textbf{D1} \nonumber \\ & \quad - 64 \xi _1^2 \xi _2^2 \sinh ^3\left(\frac{\alpha  \phi }{2}\right) (\text{sech}(\alpha  \phi )-1) \text{sech}^2(\alpha  \phi ) \text{sech}^4\left(\xi _2 \phi \right) \times \textbf{B1}\times \textbf{C} \times \Big(\alpha  \cosh \left(\frac{\alpha  \phi }{2}\right) \text{sech}(\alpha  \phi )-2 \xi _2 \sinh \left(\frac{\alpha  \phi }{2}\right) \nonumber \\ & \quad \tanh \left(\xi _2 \phi \right) \Big) \times \textbf{D1} + (\text{sech}(\alpha  \phi )-1) \text{sech}(\alpha  \phi ) \times \textbf{A1} \times \textbf{C1}  \times \bigg( 3 \alpha ^2 \text{sech}(\alpha  \phi )+32 \xi _1 \xi _2 \sinh ^3\left(\frac{\alpha  \phi }{2}\right) \text{sech}^2\left(\xi _2 \phi \right)\times  \nonumber \\ & \quad   \left(\alpha  \cosh ^3\left(\frac{\alpha  \phi }{2}\right) \text{sech}(\alpha  \phi )-\xi _2 \sinh \left(\frac{\alpha  \phi }{2}\right) \tanh \left(\xi _2 \phi \right)\right) \bigg)\times \textbf{D1} + 8 \xi _1 \xi _2 (\text{sech}(\alpha  \phi )-1) \text{sech}(\alpha  \phi ) \text{sech}^2\left(\xi _2 \phi \right) \times \textbf{A1} \times \textbf{B} \times \nonumber \\ & \quad \bigg(  -16 \xi _1 \xi _2 \sinh ^3\left(\frac{\alpha  \phi }{2}\right) \text{sech}(\alpha  \phi ) \text{sech}^2\left(\xi _2 \phi \right)  \left( \alpha  \cosh \left(\frac{\alpha  \phi }{2}\right) \text{sech}(\alpha  \phi )-2 \xi _2 \sinh \left(\frac{\alpha  \phi }{2}\right) \tanh \left(\xi _2 \phi \right)  \right) + \nonumber \\ & \quad 3 \alpha \left(  \alpha  \left(\text{sech}^2(\alpha  \phi )-\tanh ^2(\alpha  \phi )\right)-2 \xi _2 \tanh (\alpha  \phi ) \tanh \left(\xi _2 \phi \right)  \right)\bigg) \times \textbf{D1}\Bigg) \Bigg] \Bigg/ \Bigg[ \textbf{A1}^2 \times \textbf{B1} \times \textbf{C1}^3 \times \textbf{D1}  \Bigg]
\end{align}

Where, 
\begin{align*}
    \textbf{A1}=& \Big(3+ 32 \xi _1^2 \xi _2^2 \sinh ^4\left(\frac{\alpha  \phi }{2}\right) \text{sech}^2(\alpha  \phi ) \text{sech}^4\left(\xi _2 \phi \right) \Big)
\end{align*}

\begin{align*}
    \textbf{B1}=& \Big( 16 \xi _1 \xi _2 \sinh ^4\left(\frac{\alpha  \phi }{2}\right) \text{sech}(\alpha  \phi ) \text{sech}^2\left(\xi _2 \phi \right)+3 \alpha  \tanh (\alpha  \phi )  \Big) 
\end{align*}

\begin{align}
    \textbf{C1}=& \bigg(9+4 \xi _1 \xi _2 \text{sech}(\alpha  \phi ) \text{sech}^2\left(\xi _2 \phi \right) \left(8 \xi _1 \xi _2 \sinh ^4\left(\frac{\alpha  \phi }{2}\right) \text{sech}(\alpha  \phi ) \text{sech}^2\left(\xi _2 \phi \right)-3 \alpha  \tanh (\alpha  \phi )\right)\bigg) \nonumber
\end{align}

\begin{align*}
    \textbf{D1}=& \bigg(2048 \alpha  \xi _1^4 \xi _2^4 \sinh ^9\left(\frac{\alpha  \phi }{2}\right) \cosh \left(\frac{\alpha  \phi }{2}\right) \text{sech}^5(\alpha  \phi ) \text{sech}^8\left(\xi _2 \phi \right)+27 \alpha  \tanh (\alpha  \phi ) + \nonumber \\ & \quad 1536 \xi _1^2 \xi _2^3 \sinh ^6\left(\frac{\alpha  \phi }{2}\right) \text{sech}^2(\alpha  \phi ) \tanh \left(\xi _2 \phi \right) \text{sech}^4\left(\xi _2 \phi \right)+ 768 \alpha  \xi _1^3 \xi _2^3 \sinh ^6\left(\frac{\alpha  \phi }{2}\right) \text{sech}^3(\alpha  \phi ) \text{sech}^6\left(\xi _2 \phi \right) \times \nonumber \\ & \quad \left( \alpha  \left(\tanh ^2(\alpha  \phi )+\text{sech}(\alpha  \phi )\right)-2 \xi _2 \tanh (\alpha  \phi ) \tanh \left(\xi _2 \phi \right)  \right) + 36\alpha \text{sech}(\alpha  \phi )\text{sech}^2\left(\xi _2 \phi \right)\sinh ^2\left(\frac{\alpha  \phi }{2}\right)\xi _1\xi _2  \times \nonumber \\ & \quad \Big( 4 \xi _2 \tanh (\alpha  \phi ) \tanh \left(\xi _2 \phi \right)+\alpha  \text{sech}(\alpha  \phi ) ((\cosh (2 \alpha  \phi )-5) \text{sech}(\alpha  \phi )-2)  \Big) \bigg)
\end{align*}

\subsection{\underline{New Slow Roll Approximation II}}

\label{app2}
The slow-roll parameters in the case of slow-roll approximation II are given below:

\begin{align}
 \delta_1=   &\frac{4 \, \xi_1 \, \xi_2 \, \text{Sech}(\alpha \, \phi) \, \text{Sech}(\xi_2 \, \phi)^2 
    \left(16 \, \xi_1 \, \xi_2 \, \text{Sech}(\alpha \, \phi) \, \text{Sech}(\xi_2 \, \phi)^2 
    \, \text{Sinh}\left(\frac{\alpha \, \phi}{2}\right)^4 + 3 \, \alpha \, \text{Tanh}(\alpha \, \phi)\right)}
    {-9 + 18 \, \alpha \, \xi_1 \, \xi_2 \, \text{Sech}(\alpha \, \phi) \, \text{Sech}(\xi_2 \, \phi)^2 \, \text{Tanh}(\alpha \, \phi)}
\end{align}

\begin{align}
    \epsilon_1 & = \frac{-1}{{\bf{A_2}}} \bigg\{ \text{sech}(\alpha \phi) \left(16 \xi_1 \xi_2 \sinh^4\left(\frac{\alpha \phi}{2}\right) \text{sech}(\alpha \phi) \text{sech}^2(\xi_2 \phi) + 3\alpha \tanh(\alpha \phi)\right) \nonumber \\
    & \times \bigg( 32 \alpha^2 \xi_1^3 \xi_2^3 \text{sech}^6(\alpha \phi) \text{sech}^6(\xi_2 \phi) - 128 \alpha^2 \xi_1^3 \xi_2^3 \text{sech}^5(\alpha \phi) \text{sech}^6(\xi_2 \phi) + 64 \xi_1^2 \xi_2^3 \tanh(\xi_2 \phi) \text{sech}^4(\xi_2 \phi) \nonumber \\
    & -4 \xi_1 \xi_2 \text{sech}^2(\alpha \phi) \text{sech}^2(\xi_2 \phi) \bigg( 16 \alpha \xi_1^2 \xi_2^2 \tanh(\alpha \phi) \text{sech}^4(\xi_2 \phi) (\alpha \tanh(\alpha \phi) - 3 \xi_2 \tanh(\xi_2 \phi)) \nonumber\\
    & + 3 \alpha \tanh(\alpha \phi) (3 \alpha \tanh(\alpha \phi) + 2 \xi_2 \tanh(\xi_2 \phi)) - 8 \xi_1 \xi_2 \text{sech}^2(\xi_2 \phi) (\alpha \tanh(\alpha \phi) + 6 \xi_2 \tanh(\xi_2 \phi))\bigg) \nonumber\\ 
    & -4 \xi_1 \xi_2 \text{sech}^3(\alpha \phi) \text{sech}^2(\xi_2 \phi) \bigg( 3 \alpha^2 + 48 \alpha \xi_1^2 \xi_2^3 \tanh(\alpha \phi) \tanh(\xi_2 \phi) \text{sech}^4(\xi_2 \phi) + \xi_1 \xi_2 \text{sech}^2(\xi_2 \phi) \nonumber \\
    & - 3 \alpha^3 \tanh^3(\alpha \phi) + 4 \alpha \tanh(\alpha \phi) + 16 \xi_2 \tanh(\xi_2 \phi)\bigg) \nonumber \\
    & + \text{sech}(\alpha \phi) \bigg( 4 \alpha^2 \xi_1 \xi_2 \tanh^2(\alpha \phi) \text{sech}^2(\xi_2 \phi) 
\left(8 \xi_1^2 \xi_2^2 \text{sech}^4(\xi_2 \phi) + 3\right) \nonumber \\
    & + \alpha \tanh(\alpha \phi) \left(-64 \xi_1^3 \xi_2^4 \tanh(\xi_2 \phi) \text{sech}^6(\xi_2 \phi) 
-16 \xi_1^2 \xi_2^2 \text{sech}^4(\xi_2 \phi) + 24 \xi_1 \xi_2^2 \tanh(\xi_2 \phi) \text{sech}^2(\xi_2 \phi) + 9 \right) \nonumber \\
    & - 192 \xi_1^2 \xi_2^3 \tanh(\xi_2 \phi) \text{sech}^4(\xi_2 \phi) \bigg) \nonumber \\
    & + 4 \alpha \xi_1 \xi_2 \, \text{sech}^4(\alpha \phi) \text{sech}^2(\xi_2 \phi) 
\bigg( 3 \alpha + 8 \xi_1^2 \xi_2^2 \, \text{sech}^4(\xi_2 \phi) \bigg(3 \alpha + 2 \xi_2 \tanh(\alpha \phi) \tanh(\xi_2 \phi) \bigg) \bigg) \bigg)\bigg\}
\end{align}

where ${\bf{A_2}}=54 (\text{sech}(\alpha  \phi )-1)^2 \left(2 \alpha  \text{$\xi $1} \text{$\xi $2} \tanh (\alpha  \phi ) \text{sech}(\alpha  \phi ) \text{sech}^2(\text{$\xi $2} \phi )-1\right)^3$

\begin{align}
\delta_2 &= \frac{-1}{{\bf{B_{2}}}} \bigg\{ \text{sech}(\alpha  \phi ) \left(64 \text{$\xi $1}^2 \text{$\xi $2}^2 \sinh ^4\left(\frac{\alpha  \phi }{2}\right) \text{sech}^2(\alpha  \phi ) \text{sech}^4(\text{$\xi $2} \phi )-6 \alpha  \text{$\xi $1} \text{$\xi $2} \tanh (\alpha  \phi ) \text{sech}(\alpha  \phi ) \text{sech}^2(\text{$\xi $2} \phi )+9\right)  \nonumber \\
& \bigg(-128 \alpha ^2 \text{$\xi $1}^2 \text{$\xi $2}^2 \sinh ^4\left(\frac{\alpha  \phi }{2}\right) \cosh ^2\left(\frac{\alpha  \phi }{2}\right) \text{sech}^3(\alpha  \phi ) \text{sech}^4(\text{$\xi $2} \phi ) +32 \alpha ^2 \text{$\xi $1}^2 \text{$\xi $2}^2 \sinh ^4\left(\frac{\alpha  \phi }{2}\right) \text{sech}^4(\alpha  \phi ) \text{sech}^4(\text{$\xi $2} \phi ) \nonumber \\
& +32 \text{$\xi $1} \text{$\xi $2} \sinh ^3\left(\frac{\alpha  \phi }{2}\right) \text{sech}(\alpha  \phi ) \text{sech}^2(\text{$\xi $2} \phi ) \left(\alpha  \cosh \left(\frac{\alpha  \phi }{2}\right)-\sinh \left(\frac{\alpha  \phi }{2}\right) (\alpha  \tanh (\alpha  \phi )+2 \text{$\xi $2} \tanh (\text{$\xi $2} \phi ))\right)\nonumber \\ 
&+\alpha  \text{sech}^2(\alpha  \phi ) \left(32 \text{$\xi $1}^2 \text{$\xi $2}^2 \sinh ^4\left(\frac{\alpha  \phi }{2}\right) \tanh (\alpha  \phi ) \text{sech}^4(\text{$\xi $2} \phi ) (\alpha  \tanh (\alpha  \phi )+2 \text{$\xi $2} \tanh (\text{$\xi $2} \phi ))+3 \alpha \right) \nonumber \\ 
& -3 \alpha  \tanh (\alpha  \phi ) (\alpha  \tanh (\alpha  \phi )+2 \text{$\xi $2} \tanh (\text{$\xi $2} \phi ))\bigg)\bigg\}
\end{align}
\\
where ${\bf{B_2}}=27 (\text{sech}(\alpha  \phi )-1) \left(2 \alpha  \text{$\xi $1} \text{$\xi $2} \tanh (\alpha  \phi ) \text{sech}(\alpha  \phi ) \text{sech}^2(\text{$\xi $2} \phi )-1\right)^3$ 

The analytical expressions for $\epsilon_1$ and $\epsilon_2$ are given as ;

Expression for $\epsilon_2$;

\begin{align}
\epsilon_2= \frac{-1}{B_2}& 3 \, \text{Sech}[\alpha \phi] \left( -1 + 
2 \alpha \xi_1 \xi_2 \, \text{Sech}[\alpha \phi] \, \text{Sech}[\xi_2 \phi]^2 \, \text{Tanh}[\alpha \phi] \right)^3 \times \nonumber \\
& \quad \bigg( -9 - 
64 \xi_1^2 \xi_2^2 \, \text{Sech}[\alpha \phi]^2 \, \text{Sech}[\xi_2 \phi]^4 \, \text{Sinh} \left(\frac{\alpha \phi}{2} \right)^4 + 
6 \alpha \xi_1 \xi_2 \, \text{Sech}[\alpha \phi] \, \text{Sech}[\xi_2 \phi]^2 \, \text{Tanh}[\alpha \phi] \bigg) \times \nonumber \\
& \quad \bigg( \alpha (-1 + \text{Sech}[\alpha \phi]) \, \text{Tanh}[\alpha \phi] \bigg( 16 \xi_1 \xi_2 \, \text{Sech}[\alpha \phi] \, \text{Sech}[\xi_2 \phi]^2 \, \text{Sinh} \left( \frac{\alpha \phi}{2} \right)^4 + 3 \alpha \, \text{Tanh}[\alpha \phi] \bigg) \times \nonumber \\
& \quad \quad \bigg( -1 + 2 \alpha \xi_1 \xi_2 \, \text{Sech}[\alpha \phi] \, \text{Sech}[\xi_2 \phi]^2 \, \text{Tanh}[\alpha \phi] \bigg) \times \nonumber \\
& \quad \quad \bigg( -128 \alpha^2 \xi_1^3 \xi_2^3 \, \text{Sech}[\alpha \phi]^5 \, \text{Sech}[\xi_2 \phi]^6 + 
32 \alpha^2 \xi_1^3 \xi_2^3 \, \text{Sech}[\alpha \phi]^6 \, \text{Sech}[\xi_2 \phi]^6 + \nonumber \\
& \quad \quad \quad 64 \xi_1^2 \xi_2^3 \, \text{Sech}[\xi_2 \phi]^4 \, \text{Tanh}[\xi_2 \phi] - 
4 \xi_1 \xi_2 \, \text{Sech}[\alpha \phi]^2 \, \text{Sech}[\xi_2 \phi]^2 \bigg( {\bf{C_2}} \bigg) \nonumber \\
& \quad \quad + 4 \xi_1 \xi_2 \, \text{Sech}[\alpha \phi]^3 \, \text{Sech}[\xi_2 \phi]^2 \bigg( 3 \alpha^2 + 
48 \alpha \xi_1^2 \xi_2^3 \, \text{Sech}[\xi_2 \phi]^4 \, \text{Tanh}[\alpha \phi] \, \text{Tanh}[\xi_2 \phi] \nonumber \\
& \quad \quad \quad \quad + \xi_1 \xi_2 \, \text{Sech}[\xi_2 \phi]^2 \bigg( 4 \alpha \, \text{Tanh}[\alpha \phi] - 3 \alpha^3 \, \text{Tanh}[\alpha \phi]^3 + 16 \xi_2 \, \text{Tanh}[\xi_2 \phi] \bigg) \bigg) \bigg)\nonumber \\
& - 2 \alpha \, \text{Sech}[\alpha \phi] \, \text{Tanh}[\alpha \phi] \bigg( 16 \xi_1 \xi_2 \, \text{Sech}[\alpha \phi] \, \text{Sech}[\xi_2 \phi]^2 \, \text{Sinh} \left( \frac{\alpha \phi}{2} \right)^4 + 3 \alpha \, \text{Tanh}[\alpha \phi] \bigg) \times \nonumber \\
& \quad \bigg( -1 + 2 \alpha \xi_1 \xi_2 \, \text{Sech}[\alpha \phi] \, \text{Sech}[\xi_2 \phi]^2 \, \text{Tanh}[\alpha \phi] \bigg) \times \nonumber \\
& \quad \bigg( -128 \alpha^2 \xi_1^3 \xi_2^3 \, \text{Sech}[\alpha \phi]^5 \, \text{Sech}[\xi_2 \phi]^6 + 
32 \alpha^2 \xi_1^3 \xi_2^3 \, \text{Sech}[\alpha \phi]^6 \, \text{Sech}[\xi_2 \phi]^6 + \nonumber \\
& \quad \quad \quad 64 \xi_1^2 \xi_2^3 \, \text{Sech}[\xi_2 \phi]^4 \, \text{Tanh}[\xi_2 \phi] - 
4 \xi_1 \xi_2 \, \text{Sech}[\alpha \phi]^2 \, \text{Sech}[\xi_2 \phi]^2 \bigg( {\bf{C_2}} \bigg) \nonumber \\
& \quad \quad + 4 \xi_1 \xi_2 \, \text{Sech}[\alpha \phi]^3 \, \text{Sech}[\xi_2 \phi]^2 \bigg( 3 \alpha^2 + 
48 \alpha \xi_1^2 \xi_2^3 \, \text{Sech}[\xi_2 \phi]^4 \, \text{Tanh}[\alpha \phi] \, \text{Tanh}[\xi_2 \phi] \nonumber \\
& \quad \quad \quad \quad + \xi_1 \xi_2 \, \text{Sech}[\xi_2 \phi]^2 \bigg( 4 \alpha \, \text{Tanh}[\alpha \phi] - 3 \alpha^3 \, \text{Tanh}[\alpha \phi]^3 + 16 \xi_2 \, \text{Tanh}[\xi_2 \phi] \bigg) \bigg) \bigg)
\end{align}
where ${\bf{B_2}}$ is given as;

\begin{align*}
& \left( (1 - \text{Sech}[\alpha \phi]) \left( -1 + \text{Sech}[\alpha \phi] \right) \left( 9 - 
18 \alpha \xi_1 \xi_2 \, \text{Sech}[\alpha \phi] \, \text{Sech}[\xi_2 \phi]^2 \, \text{Tanh}[\alpha \phi] \right)^2 \right. \\
& \quad \left. \times \left( 1 - 
2 \alpha \xi_1 \xi_2 \, \text{Sech}[\alpha \phi] \, \text{Sech}[\xi_2 \phi]^2 \, \text{Tanh}[\alpha \phi] \right)^4 \times \right. \\
& \quad \left. \bigg( -128 \alpha^2 \xi_1^3 \xi_2^3 \, \text{Sech}[\alpha \phi]^5 \, \text{Sech}[\xi_2 \phi]^6 + 
32 \alpha^2 \xi_1^3 \xi_2^3 \, \text{Sech}[\alpha \phi]^6 \, \text{Sech}[\xi_2 \phi]^6 + \right. \\
& \quad \left. 64 \xi_1^2 \xi_2^3 \, \text{Sech}[\xi_2 \phi]^4 \, \text{Tanh}[\xi_2 \phi] - 
4 \xi_1 \xi_2 \, \text{Sech}[\alpha \phi]^2 \, \text{Sech}[\xi_2 \phi]^2 \bigg( {\bf{C_2}} \bigg) \right. \\
& \quad \left. - 4 \xi_1 \xi_2 \, \text{Sech}[\alpha \phi]^3 \, \text{Sech}[\xi_2 \phi]^2 \bigg( 3 \alpha^2 + 
48 \alpha \xi_1^2 \xi_2^3 \, \text{Sech}[\xi_2 \phi]^4 \, \text{Tanh}[\alpha \phi] \, \text{Tanh}[\xi_2 \phi] \right. \\
& \quad \left. + \xi_1 \xi_2 \, \text{Sech}[\xi_2 \phi]^2 \bigg( 4 \alpha \, \text{Tanh}[\alpha \phi] - 
3 \alpha^3 \, \text{Tanh}[\alpha \phi]^3 + 
16 \xi_2 \, \text{Tanh}[\xi_2 \phi] \bigg) \bigg) \right. \\
& \quad \left. + \text{Sech}[\alpha \phi] \bigg( 4 \alpha^2 \xi_1 \xi_2 \, \text{Sech}[\xi_2 \phi]^2 \left( 3 + 
8 \xi_1^2 \xi_2^2 \, \text{Sech}[\xi_2 \phi]^4 \right) \, \text{Tanh}[\alpha \phi]^2 \right. \\
& \quad \left. - 192 \xi_1^2 \xi_2^3 \, \text{Sech}[\xi_2 \phi]^4 \, \text{Tanh}[\xi_2 \phi] + 
\alpha \, \text{Tanh}[\alpha \phi] \left( 9 - 
16 \xi_1^2 \xi_2^2 \, \text{Sech}[\xi_2 \phi]^4 + 
24 \xi_1 \xi_2^2 \, \text{Sech}[\xi_2 \phi]^2 \, \text{Tanh}[\xi_2 \phi] \right. \right. \\
& \quad \left. \left. - 64 \xi_1^3 \xi_2^4 \, \text{Sech}[\xi_2 \phi]^6 \, \text{Tanh}[\xi_2 \phi] \right) \bigg) \right. \\
& \quad \left. + 4 \alpha \xi_1 \xi_2 \, \text{Sech}[\alpha \phi]^4 \, \text{Sech}[\xi_2 \phi]^2 \bigg( 3 \alpha + 
8 \xi_1^2 \xi_2^2 \, \text{Sech}[\xi_2 \phi]^4 \left( 3 \alpha + 
2 \xi_2 \, \text{Tanh}[\alpha \phi] \, \text{Tanh}[\xi_2 \phi] \right) \bigg) \right.
\end{align*}

${\bf{C_2}}=16 \alpha \xi_1^2 \xi_2^2 \, \text{Sech}[\xi_2 \phi]^4 \, \text{Tanh}[\alpha \phi] \bigg( \alpha \, \text{Tanh}[\alpha \phi] - 3 \xi_2 \, \text{Tanh}[\xi_2 \phi] \bigg) \nonumber \\
 \quad \quad \quad \quad + 3 \alpha \; \text{Tanh} [\alpha \phi] \bigg( 3 \alpha \;  \text{Tanh}[\alpha \phi] + 2 \xi_2 \; \text{Tanh}[\xi_2 \phi] \bigg) - 8 \xi_1 \xi_2 \, \text{Sech}[\xi_2 \phi]^2 \bigg( \alpha \; \text{Tanh}[\alpha \phi] + 6 \xi_2 \; \text{Tanh}[\xi_2 \phi] \bigg)$

\end{document}